\documentclass[12pt]{article}
\usepackage{latexsym,epsfig,graphicx,amsmath,amssymb,amscd,multirow, microtype,
setspace,psfrag,paralist,dsfont,amssymb, multicol}

\usepackage{appendix}
\usepackage{caption}
\usepackage{subcaption}
\usepackage{hhline}
\usepackage{epstopdf}
\usepackage{bm}

\usepackage[round]{natbib}

\usepackage[british]{babel}

\usepackage{array}
\newcolumntype{P}[1]{>{\centering\arraybackslash}p{#1}}
\usepackage[table]{xcolor}
\usepackage{siunitx,booktabs}
\usepackage{diagbox}

\newcommand{\thetavec}{{\boldsymbol{\theta}}}

\newcommand{\Sigmavec}{{\boldsymbol{\Sigma}}}

\newcommand{\yvec}{{\boldsymbol{y}}}

\newcommand{\Avec}{{\boldsymbol{A}}}
\newcommand{\Cvec}{{\mathbf{C}}}
\newcommand{\Bvec}{{\mathbf{B}}}
\newcommand{\Wvec}{{\boldsymbol{W}}}
\newcommand{\vvec}{{\boldsymbol{v}}}
\newcommand{\avec}{{\boldsymbol{a}}}

\newcommand{\pr}{{\rm Pr}}

\newcommand{\Var}{{\rm Var}}

\newcommand{\thetavechat}{\widehat{\thetavec}}

\newcommand{\Xvec}{\boldsymbol{X}}

\newcommand{\xvec}{\boldsymbol{x}}
\newcommand{\muvec}{\boldsymbol{\mu}}

\newcommand{\F}{\mathcal{F}}

\newcommand{\ffrac}[2]{\ensuremath{\frac{\displaystyle #1}{\displaystyle #2}}}

\begin{document}

\pagenumbering{gobble}
\clearpage
\thispagestyle{empty}

\title{Modeling Bivariate Geyser Eruption System with Covariate-Adjusted Recurrent Event Process}

\author{Zhongnan Jin$^1$, Lu Lu$^2$, Khaled Bedair$^3$, and Yili Hong$^1$\\[2ex]
{\small$^1$Department of Statistics, Virginia Tech, Blacksburg, VA, USA}\\
{\small $^2$Department of Mathematics \& Statistics, University of South Florida, Tampa, FL, USA}\\
{\small $^{3}$Department of Statistics \& Mathematics, Faculty of Commerce, Tanta University, Egypt}
}

\date{}

\maketitle

\begin{abstract}
Geyser eruption is one of the most popular signature attractions at the Yellowstone National Park. The interdependence of geyser eruptions and impacts of covariates are of interest to researchers in geyser studies. In this paper, we propose a parametric covariate-adjusted recurrent event model for estimating the eruption gap time. We describe a general bivariate recurrent event process, where a bivariate lognormal distribution  and  a Gumbel copula with different  marginal distributions are used to model an interdependent dual-type event system. The maximum likelihood approach is used to estimate model parameters. The proposed method is applied to analyzing the Yellowstone geyser eruption data for a bivariate geyser system and offers a deeper understanding of the event occurrence mechanism of individual events as well as the system as a whole. A comprehensive simulation study is conducted to evaluate the performance of the proposed method.


\end{abstract}


\newpage

\pagenumbering{arabic}

\section{Introduction}
\subsection{Background}
Geyser eruption is one of the signature attractions at the Yellowstone National Park, which is home to two-thirds of the worlds' geysers. Tourists around the world crave to witness this fascinating natural phenomenon. Many researchers are interested in studying geyser eruptions and the underlying mechanisms. \citet{Fournier1969} built a physical model to describe the time interval between eruptions for the Old Faithful geyser, which is one of the most famous geysers in the Yellowstone National Park. \citet{Rinehart1972} showed that the Old Faithful Geyser activities are affected by earth tidal forces, barometric pressure, and tectonic stresses. However, geyser eruptions have not been studied by statistical methods. This paper develops a statistical model for analyzing geyser eruption data from a bivariate geyser system. This work will benefit the geyser study community for understanding and effectively modeling geyser eruption activities.

Typical geyser eruption is a repeating process and hence can be modeled with a recurrent process for events repeatedly occurring over time. Recurrent processes have had broad applications in diverse areas. For example, they have been widely used for studying vehicle failures in warranty studies \citep{Lawless1995}, relapse biomarkers in cancer research \citep{Schaubel2004}, and sports injury analysis \citep{Ullah2014}. Typically the time interval between two consecutive events, which is also referred to as the gap time, is studied to model the event frequency in a recurrent process. The proportional intensity  models (\citealp{Cox1972}, and \citealp{AndersenGill1982}) are popular for modeling event occurrences of a single type of event. In more sophisticated studies, there are multiple types of recurrent events observed in a single system. The occurrence of any type of event will result in a system event. In addition, in a multi-type recurrent event process, the gap time for different event types could be correlated. For example, the occurrence of one type of event could cause other types of events to occur more frequently. In this case, a multivariate recurrent process should be considered to model the interdependence of multiple event types in the same system.

In many event analyses, covariates are found to be useful for modeling the event occurrence time and frequency. Many recurrent processes are affected by process conditions. For example, some mechanical failures could occur at a higher frequency under a higher temperature, humidity, or pressure. Incorporating covariates into the recurrent process models could improve the model performance and provide a more precise estimation of the event time and frequency. Models of this type are referred to as the covariate-adjusted recurrent event process models.

In this paper, we focus on modeling and analysis of geyser eruptions for a two geyser system in the Yellowstone National Park. In a multi-type recurrent event system, the system events can result from either type of events, and hence any consecutive events could be associated with the same or different event types. In order to describe this bivariate recurrent event process, we need to not only understand the marginal behavior of each type of event, but also understand the interdependence between the two types of events. We consider a bivariate distribution for the gap times between successive events for a bivariate event system. To improve the estimation, we leverage the covariate information on the eruption duration and develop a covariate-adjusted bivariate recurrent event process model for estimating the eruption gap time of a two geyser system formed by West Triplet Geyser and Grotto Geyser in the Yellowstone National Park.

\subsection{Related Literature}


Recurrent event processes are extensively studied in the areas of reliability, public health, and medical studies. The nonparametric estimation of gap time distribution based on multivariate failure time data was introduced in \citet{Schaubel2004}. \citet{Dauxois2009} considered the risks of two nosocomial infections for patients admitted to hospitals. \citet{Bouaziz2013} provided a nonparametric method to estimate the intensity function of a recurrent process. Other than hazard functions and intensity functions, survival status in time is often of interest as well. \citet{Huang2007} investigated the disease free survival rate in a recurrent heart failure study. \citet{Zeng2009} and \citet{Garre2008} focused on studying the terminal events in recurrent systems. \citet{meyer2015bayesian} presented Bayesian analysis of recurrent event using copulas. An earlier review regarding recurrent events can be found in \citet{Lawless1995} and a review on modeling of repairable systems can be found in \citet{Lindqvist2006}. Classical books on recurrent event data analysis include \citet{vere2003introduction}, \citet{CookLawless2007}, and \citet{DuchateauJanssen2008}.

In medical research, \citet{Liu2009} presented repeated measurements of biomarker  to  determine the HIV survival status in  a recurrent event system. \citet{Sun2006} applied covariate-adjusted additive hazard model for the data, which is involving recurrent gap times. \citet{PrasadRao2002} used a proportional hazard function with covariate adjustment in a repairable system. In another application of recurrent event data, \citet{Huzurbazar2010} incorporated covariates in  a flowgraph model. \citet{YangZhangHong2013} introduced multivariate lognormal assumption on event gap times  of different event types. \citet{YangHongZhangLi2017} considered a parametric model for the multi-type event recurrent event data without covariates and developed copula function on gap times for the recurrent process in a car body manufacturing process.

Existing research does not consider correlated renewal process with covariate adjustments. Motivated by the geyser data, we propose the CARP model to study the eruption gap time for a two-geyser system. Application-wise, geyser eruption is rarely studied  by statistical models. For the geyser eruption study, the modeling and analysis presented in this paper are new to geyser research.

\subsection{Overview}
The rest of this paper is organized as follows. Section \ref{sec: geyser data} provides more details on the geyser eruption data from the Yellowstone National Park. Section \ref{sec:data.and.model} discusses the model formulation for the multi-type recurrent event system. Section \ref{sec:par.est} describes the maximum likelihood approach for estimating the model parameters. A simulation study is described in Section~\ref{sec:simulation} to evaluate the proposed method with model comparisons made under different parameter settings. The modeling and analysis of the Yellowstone geyser data are detailed in Section~\ref{sec:application}. Section~\ref{sec:conclusion} contains some concluding remarks.

\section{Geyser Data}\label{sec: geyser data}

We use the publicly available Yellowstone geyser eruption data, which were collected in 2008 by the Geyser Observation and Study Association (GOSA). By using underground sensors, water levels were measured continuously, and occurrences of geyser eruptions were detected automatically. For each geyser, the GOSA data include the starting time and duration of each eruption. We choose to analyze the data from the West Triplet Geyser and the Grotto Geyser during the study period between June 2008 and November 2008. This particular dataset and study period were chosen to ensure that completely uninterrupted recurrent event data are available over a relatively long time span to allow for the dependence modeling for the two geysers.

We merge their eruption records in a temporal order as illustrated in Table~\ref{table:data.example}. The data include the date and time when an eruption occurred, the eruption duration, and which geyser had the eruption. During the study period, the West Triplet Geyser erupted more frequently than the Grotto Geyser. Also, the eruptions of the West Triplet Geyser also lasted longer than those of the Grotto Geyser on average.  More specifically, the average eruption gap time for the West Triplet Geyser is 6.8 hours with a standard deviation of 2.8 hours, while the Grotto Geyser had an average eruption gap time of 9.3 hours with a standard deviation of 8.6 hours. For the eruption duration, the Grotto Geyser eruptions lasted generally longer than the West Triplet Geyser, with the average duration of the Grotto Geyser eruptions being 3.5 hours with a standard deviation of 5.2 hours and the eruption duration of the West Triplet Geyser averaged at 0.7 hours with a 0.5 hours standard deviation.

\begin{figure}
	\centering
	\begin{subfigure}{0.48\textwidth}
		\centering
        \includegraphics[width=\textwidth]{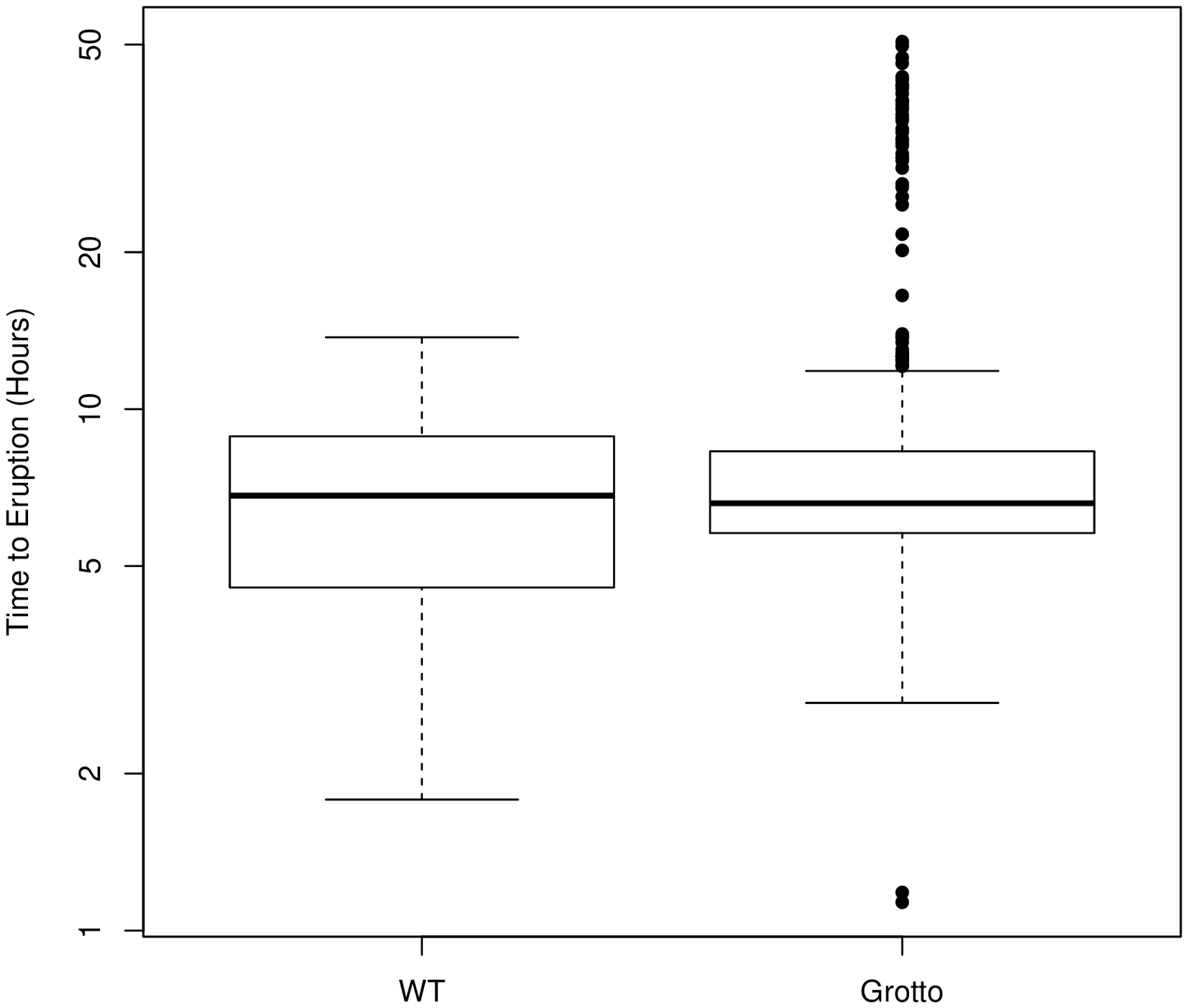}		
        \caption{Time to eruption (Hour)}
		\label{Fig:time.to.eruption}
	\end{subfigure}%
	~
	\begin{subfigure}{0.48\textwidth}
		\centering
        \includegraphics[width=\textwidth]{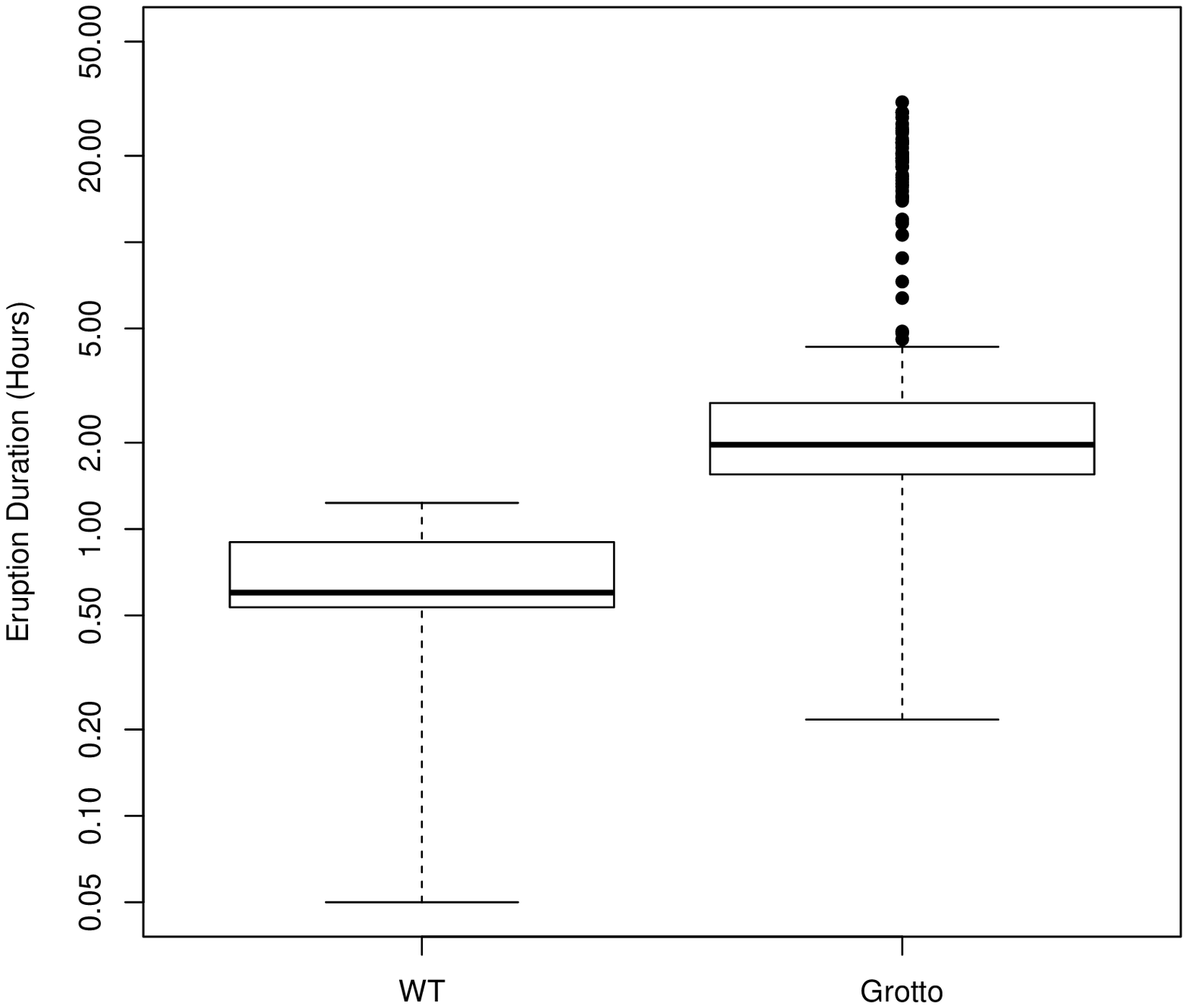}		
        \caption{Eruption duration (Hour)}
		\label{Fig:duration}
	\end{subfigure}
	\caption{Boxplots for the time to eruption and the eruption duration for the West Triplet Geyser and Grotto Geyser from June to November in 2008. Note that the $y-$axis is on log scale.}
\end{figure}

\begin{figure}
	\centering
	\begin{subfigure}{0.48\textwidth}
		\centering
		\includegraphics[width=\textwidth]{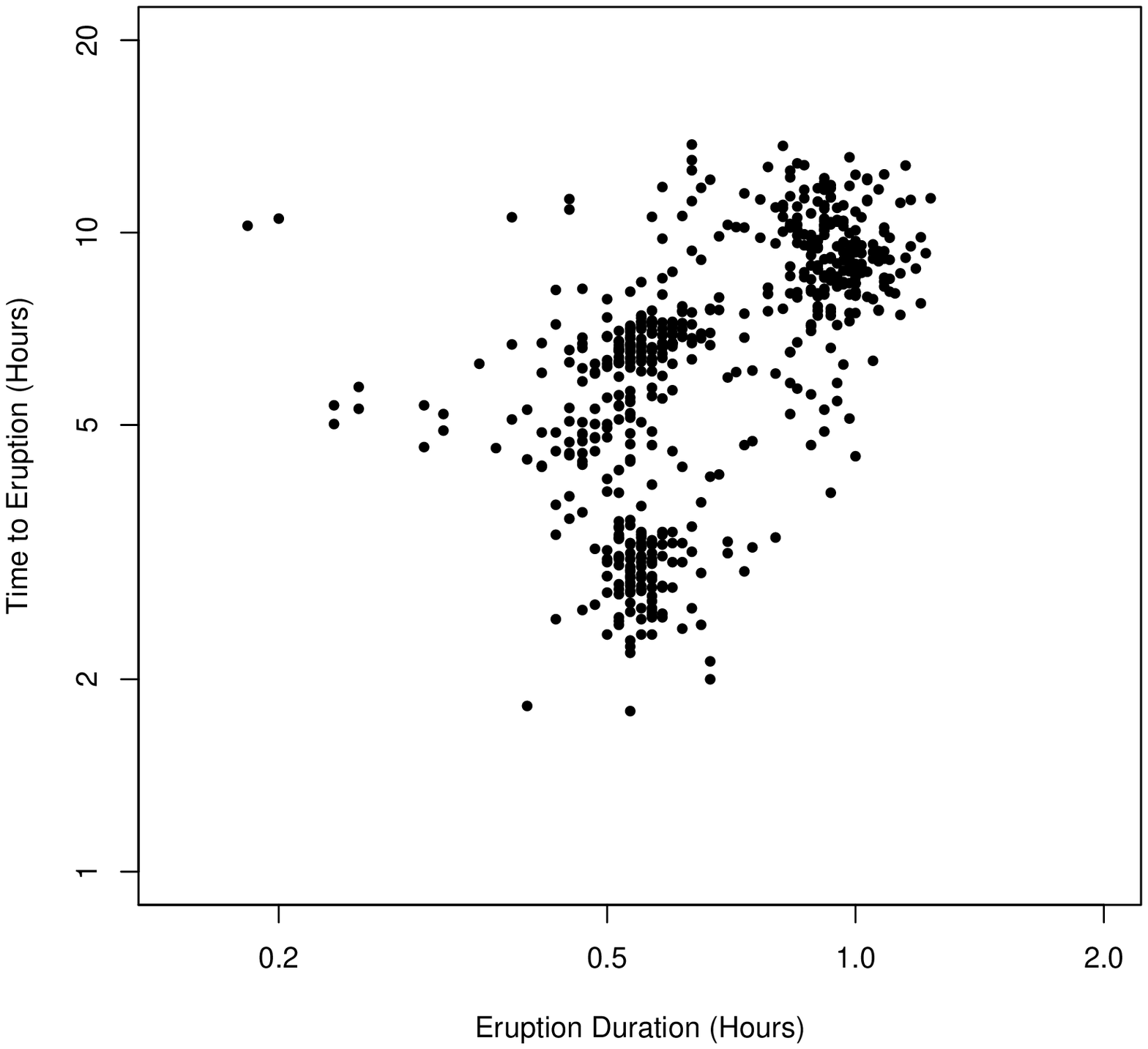}
		\caption{The West Triplet Geyser}
	
	\end{subfigure}%
	~
	\begin{subfigure}{0.48\textwidth}
		\centering
		\includegraphics[width=\textwidth]{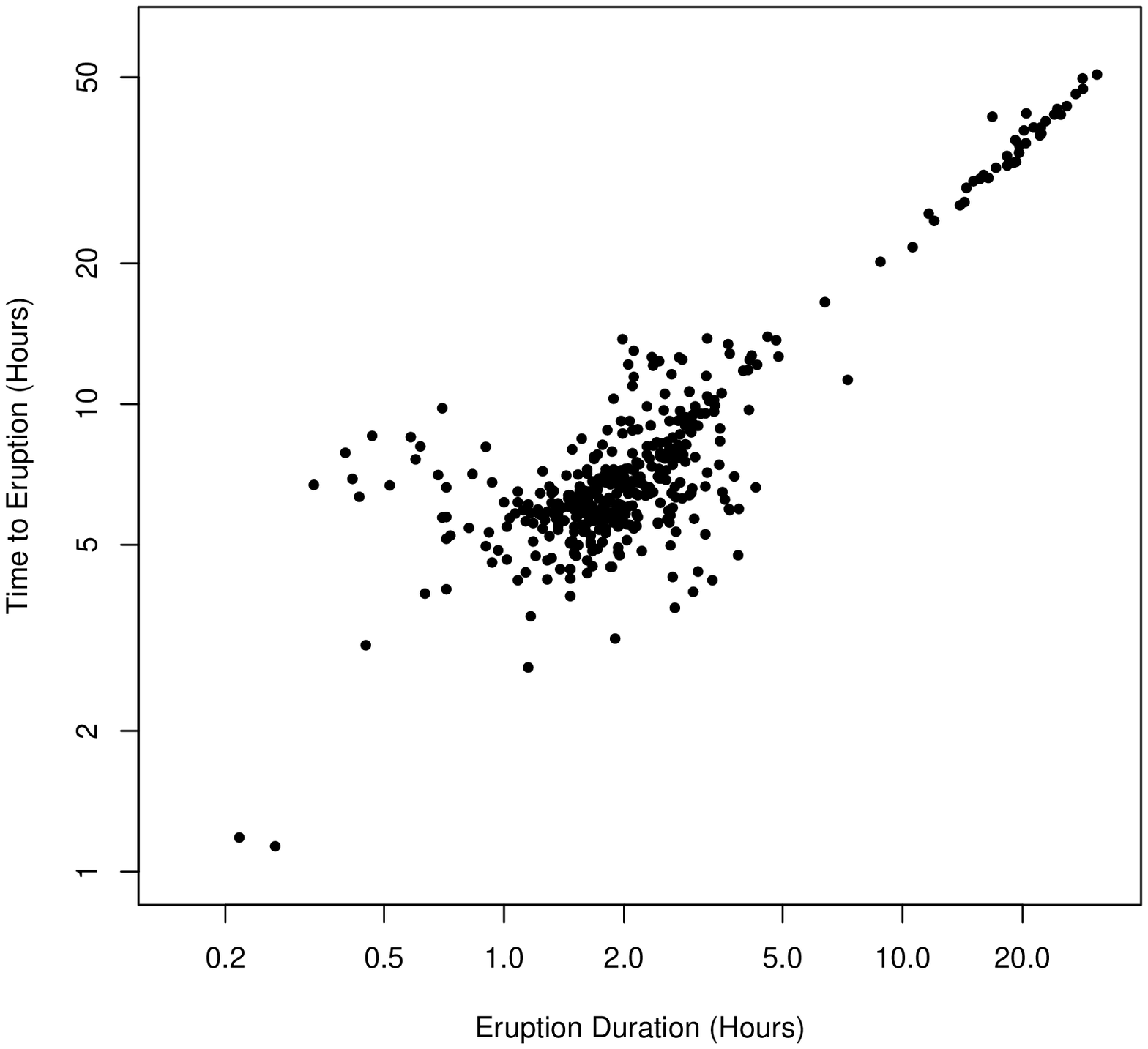}
		\caption{The Grotto Geyser}
	\end{subfigure}
	\caption{The plot of the time between eruptions versus the eruption duration for (a) the West Triplet Geyser and (b) the Grotto Geyser from June to November in 2008. Note that both the $x-$axis and the $y-$axis are on log scale.}\label{Fig:gap_vs_duration}
\end{figure}

Figures~\ref{Fig:time.to.eruption} and~\ref{Fig:duration} display the side-by-side boxplots of the time between eruptions and the duration of eruptions, respectively, for the West Triplet and Grotto geysers. We can see compared with the West Triplet Geyser, the Grotto Geyser has a much larger variation of the eruption gap time with a number of extremely long gaps between eruptions and also more variation in the eruption frequency. On the other hand, the West Triplet Geyser has considerably shorter and less variable eruption durations than the Grotto Geyser. In addition,  Figure~\ref{Fig:gap_vs_duration} shows the plots of the time between eruptions versus the duration time of the previous eruption for both geysers. We can observe a moderate correlation between these two variables for the West Triplet Geyser and a strong correlation for the Grotto geyser. Hence, we decided to utilize the duration of the previous eruption for both geysers to help model the eruption gap time.

\begin{figure}
	\centering
	\includegraphics[width=0.8\textwidth]{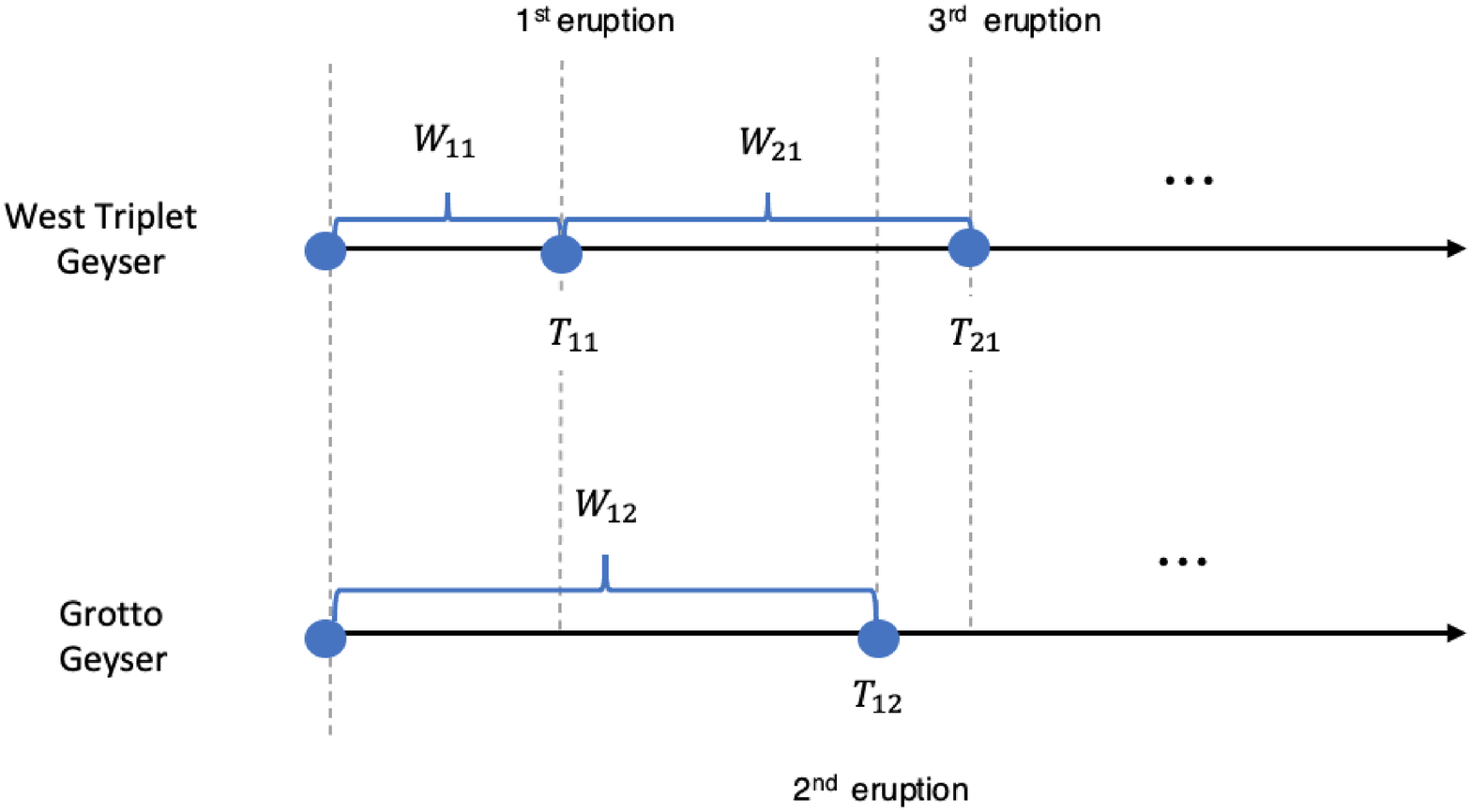}
	\caption{Illustration of geyser eruptions with West Triplet and Grotto Geysers. }
	\label{fig:Geyser_plot}
\end{figure}

Figure \ref{fig:Geyser_plot} illustrates the data obtained for the two-geyser system. In this figure, $W_{kj}$ denotes the gap time of the $k$th eruption (i.e., the time interval between the $(k-1)$th and $k$th eruptions) for geyser $j$ where $j=1$ for the West Triplet Geyser and $j=2$ for the Grotto Geyser. The covariate $x_{i}$ denotes the eruption duration for the $i$th eruption in the bivariate geyser system, where $i=1,\cdots,n$, $n=n_1+n_2$ is the total number of eruptions for both geysers, and $n_j$ denotes the number of eruptions for the $j$th geyser. In particular, $n_1=580$, $n_2=421$, and $n=1001$ for the dataset we analyzed. Note here the time between eruptions are labeled separately for each individual geyser. To model the bivariate geyser system, we will introduce new notation for jointly describing the event time, event type and covariate information in Section~\ref{sec:data.and.model}.

\begin{table}
	\caption{Sample observations from the West Triplet and Grotto Geysers.}\label{table:data.example}
	\begin{center}
		\begin{tabular}{  c | c| c }
			\hline
			\hline
			Eruption time & Duration (hours) & Geyser\\
			\hline
			2008-06-20 16:58:00 & 0.93 & Grotto \\
			\hline
			2008-06-20 20:46:00 & 0.75 & West Triplet \\
			\hline
			2008-06-20 21:31:00 & 2.05 & Grotto \\
			\hline
			2008-06-21 02:51:00 & 1.08  &   West Triplet \\
			\hline
			2008-06-21 04:48:00 & 2.63 & Grotto\\
			\hline
			2008-06-21 11:15:00 & 1.72 & Grotto\\
			\hline
			2008-06-21 13:02:00 & 0.48  &  West Triplet\\
			\hline
			2008-06-21 17:56:00 & 3.58 & Grotto\\
			\hline
			2008-06-21 18:20:00 & 0.73  & West Triplet\\
			\hline
			2008-06-22 00:11:00 & 0.93  & Grotto\\
			\hline
			\hline
		\end{tabular}
	\end{center}
	
\end{table}

\section{Data Setup and Model}\label{sec:data.and.model}

\subsection{Data Setup}
\label{subsection:proposed.model}

Suppose that in a bivariate recurrent process with $n$ total events, the systematic event time is described by the variable $T_i$ for $i=1,\cdots,n$. We use $T_{0}$ to denote the starting time or the system installation time, and $T_n$ denotes the last event time in the system. In a bivariate system, there are two types of events, and hence we use an indicator variable $\Delta_{i} \in \{ 1,2\}$ to represent the type of event. For an event that occurs at time $t_i$, the covariate vector on the event duration is denoted by $\Xvec_{i}$, which is a vector of the previous duration for both types of events. Therefore, each event can be represented by $\{T_{i}, \Delta_{i}, \Xvec_{i}\}$, where $i = 1, \cdots, n$. We use $\{t_{i}, \delta_{i}, \xvec_{i}\}$ to denote the observations of the $i$th event which occurs at time $t_{i}$ and it is from event type $\delta_{i}$ with covariates measured as $\xvec_{i}$. Note in an observed recurrent process with $n$ total events, the last event is observed at time $t_{n}$, where $t_{n}\leq \kappa$ with $\kappa$ being the pre-defined study termination time.

This sequence of bivariate events can be expressed as a counting process $\{N(t):t\ge 0\}$, where $N(t)$ denotes the cumulative number of events at time $t$ regardless of the event type. Similarly, we define the counting process for each individual event type $j$ as $\{N_{j}(t): t\ge 0\}$, where $j = 1$ or $2$. In addition, we denote the event history of the system up to a time point $s \leq \kappa$ by $\mathcal{H}_{s} = \{N(t): t \le s\}.$ Similarly, the covariate history can be expressed as $\mathcal{X}_{s} = \{\xvec_{t}: t \le s\}.$  For further discussion, we use $\F_{s} = \{ \mathcal{H}_{s}, \mathcal{X}_{s}  \}$ to denote the history including both event and covariate information.

For a bivariate recurrent process system, event time variable $T_i$ defined above satisfies
$$0=T_{0}<T_{1}<\cdots<T_{i}<\cdots<T_n\leq\kappa.$$
Similarly, for events of type $j$, the event time variables are defined as $T_{lj}$, where $l = 1, \dots, n_{j}$ and $n_{j}$ is the number of events from type $j$. In a bivariate recurrent system, we have the relation $n_{1} + n_{2} = n$. The time variable $T_{lj}$ is also presented in temporal order. For the events of type $j$, $ j = 1$ or $2$, we have
$$0=T_{0j}<T_{1j}<\cdots<T_{lj}<\cdots<T_{n_{j}, j}\leq\kappa.$$
Based on the ordered event times, the event gap time for the $j$th type as the interval between two consecutive events can be calculated as $W_{lj}=T_{l+1, j}-T_{lj}$, where $l=0, 1, \dots, n_{j}-1$.

Now consider the joint bivariate recurrent system. For the $i$th event in the system, the two-dimensional event gap time variable is defined as
$$\Wvec_i=(W_{l_{i1}, 1}, W_{l_{i2}, 2})',$$
where $l_{ij}=N_j(t_i)$ is  the cumulative number of events for $j$th event type by time $t_{i}$. For example, at the system installation time where $t = 0$, the event gap time vector is $\Wvec_{0} =(W_{01}, W_{02})'$. If the first event occurs at time $t_{1}$ and is from event type 1, then the event gap time vector is denoted as $\Wvec_{1} =(W_{11}, W_{02})'$. In this case, $l_{11}=1$ since one event of type~1 has occurred as of time $t_1$, while $l_{12}=0$ as no event of type~2 has occurred by that time point.

To link the event gap time $\Wvec_i$ with the system event time $T_i$, we introduce an age variable at time $t$ which is defined as the time between the time point $t$ and the time of the latest event of each event type. In a bivariate recurrent process, the age variable is a vector of two components denoted as
$$\Avec_{i}=[A_{1}(T_i), A_{2}(T_i)]',$$
where $A_{j}(t)= t -T_{N_j(t),\,j}$ for $j = 1$ and $2$. For an observed event at time $t_i$, the observed age vector is denoted as $\avec_{i}=[a_{1}(t_{i}), a_{2}(t_{i})]'$, where $a_{j}(t)=t-t_{N_j(t),\,j}$. For example, at the system installation time at $t = 0$, the age vector is $\avec_{0}=(0,0)'$. If the first event occurs from type 1 at time $t_1$, then $\avec_{1}=(0, t_1)'$. The age variable then connects between the system event time and the event gap time $\Wvec_{i}$ through the relation $\Wvec_i \geq \Avec_i$. Here, the vector comparison is defined to be an element-wise comparison. In other words, $\Wvec_i \geq \Avec_i$ suggests $W_{l_{ij}, j}\geq A_j(T_i)$ for both $j=1$ and $2$.

\subsection{Model}
In this section, we introduce the proposed covariate-adjusted recurrent process (CARP). After an event occurred at time $t_i$ (or after the system installation at time $t_0=0$), the distribution of the event gap time is given by
\begin{align}\label{eqn:cond.prp}
\Wvec_i|\F_{t_i}\sim F_{\Wvec}(\vvec|\avec_i, \xvec_{i}), \quad i=0, 1,2,\cdots, n.
\end{align}
Here, $\Wvec=(W_1, W_2)'$, $\vvec=(v_1, v_2)'$, and
\begin{align}\label{eqn:prp.renewal}
F_{\Wvec}(\vvec|\avec_i, \xvec_i)=\pr(\Wvec_i\leq\vvec|\Wvec_i\geq\avec_i, \xvec_i)
\end{align}
is the joint cumulative distribution function (cdf) of the event gap time conditioned on the age and covariates. At any event time, the information we have about the event type that has not yet occurred is captured through its age and the conditional probability that is conditioned on the event eruption time is greater than or equal to the age since the last eruption. At each event time, the age is set to be zero for the occurred event type. The event gap time variable $\Wvec_{i}$ is adjusted by covariates $\xvec_i$ which is further discussed in Section \ref{sec:Net_dependence} in more detail.

\subsection{Dependence Modeling and Covariate Adjustment}
\label{sec:Net_dependence}
Dependence between events from different types in a bivariate system is modeled by implementing distributional assumptions on variable $\Wvec_i = (W_{i1}, W_{i2})'$. In this paper, we use a bivariate lognormal distribution and a copula function to model the random vector $\Wvec_{i}$, where in both models, covariates $\xvec_i$ are used for adjustment. We refer to the CARP models under these assumptions as the CARP-MLN and CARP-copula models, respectively. Here MLN is short for multivariate lognormal.

\subsubsection*{The CARP-MLN Model}
For the multivariate lognormal distribution,
$$\Wvec_i\sim \textrm{MLN}[\muvec(\xvec_i), \Sigmavec],$$
where the location parameter in the bivariate lognormal distribution is expressed as a linear form of covariates $\xvec_{i}$. That is,
\begin{align}\label{eqn:linear_tran}
\muvec(\xvec_{i}) = \muvec_{0} + \Bvec  \xvec_{i}\,.
\end{align}
In the linear expression above, $\muvec_{0}$ is a vector of baseline location parameters and $\Bvec$ is a 2 $\times$ 2 coefficient matrix. In the bivariate lognormal assumption, we use a covariance matrix $\Sigmavec$ to capture the event dependence between events from two event types. Specifically, the diagonal elements in $\Sigmavec$ represent marginal variances while the off diagonal elements stand for the covariances. When using the bivariate lognormal distribution, the covariance matrix is defined as $\Sigmavec = \Cvec \Cvec'$ to ensure $\Sigmavec$ to be positive definite, where
\begin{align}\label{eqn:C_matrix}
\Cvec = \left(\begin{array}{cc} \sigma_{1} & 0 \\ \eta & \sigma_{2} \end{array}\right).
\end{align}
The correlation $\rho$ is determined by $\sigma_2$ and $\eta$ as
$$\rho =\frac{ \eta} {\sqrt{\sigma^2_2 + \eta^2} }.$$
Note the above covariance matrix offers great flexibility to model different correlation relationships of varied size and direction. The sign of the $\eta$ value determines if the two types of events have a positive or negative correlation. In a bivariate lognormal distribution, the marginal distribution of each dimension also follows a lognormal distribution. Therefore, when the observed marginal distributions do not seem to follow the lognormal distributions, or their dependency cannot be characterized by the covariance matrix $\Sigmavec$, the assumed bivariate lognormal distribution is not appropriate. In this case, the alternative strategy is to define $W_{1}$ and $W_{2}$ by separate distributions and combine them through a more flexible copula function. This is referred to as the CARP-copula model, which will be introduced in the next section.

\subsubsection*{The CARP-copula Model}

Let $F_{1}$ and $F_{2}$ denote the marginal cdfs of $W_1$ and $W_2$. There always exists a copula function $C$ such that the joint cdf of the two dimensional variable $(W_{1}, W_{2})'$ can be written as
\begin{align}\label{eqn:sklar}
F(v_{1}, v_{2}) = C[F_{1}(v_{1}), F_{2}(v_{2})],
\end{align}
where for any unitary uniform variable $U_{j}, j = 1, 2$, the bivariate copula is defined as
$$
C(u_{1}, u_{2}) = \pr(U_{1} \le u_{1}, U_{2} \le u_{2}).
$$
This is also known as the Sklar's Theorem. By using a copula function, we have the flexibility to choose marginal distributions separately for each event type. For instance, a Gamma distribution and a Weibull distribution can be used as marginal distributions for the two types of events, respectively. With selected $F_{1}$ and $F_{2}$, one can combine the marginal distributions by using different copula functions.

Note the copula approach allows us to model the marginal distributions separately and then combine them through an appropriate copula function for modeling the dependence structure. For the geyser system, we choose to use parametric distributions for modeling the marginal distributions. However, the method can be easily generalized to using nonparametric methods for modeling the marginal distributions and hence offers great flexibility to be adapted for broad applications.

In the literature, a variety of copula functions has been introduced to capture  different dependence patterns among the marginal distributions. A Gaussian copula uses a multivariate normal distribution of transformed marginal distributions based on the inverse cdf of the standard normal distribution. The Archimedean copulas are an associative class of copula functions that model the multivariate dependence through a single parameter. The power variance function copulas including Clayton, Gumbel and Inverse Gaussian are among the most popular ones that are flexible for modeling various dependence structures (e.g., \citealp{romeo2018bayesian}). In this paper, we use the Gumbel copula function which is popular for modeling stronger dependence in the positive tail. We refer to the CARP model with the use of Gumbel copula as CARP-copula model for the rest of the paper. In fact, the CARP-MLN model is a special case of the CARP-copula model, where in CARP-MLN, the Gaussian copula is applied and lognormal marginal distributions are selected.

The CARP-MLN model characterizes the dependence among event types through the covariance matrix $\Sigmavec$, while the CARP-copula model quantifies dependence using the copula parameter. Particularly, in the Gumbel copula model, the parameter is denoted as $\alpha$. As a result, in the CARP-MLN and CARP-copula models, we have different parameters to characterize event dependence. In order to compare dependence from different models, we introduce the Kendall's tau.

In the CARP-copula model, the Kendall's tau is expressed as $\tau = 1 -1/\alpha,$ where $\alpha$ is the copula coefficient in the Gumbel copula. In the CARP-MLN model, the Kendall's tau is calculated as $ \tau = (2/\pi)\arcsin(\eta/\sqrt{\sigma_2^2+\eta^2}),$ where $\eta$ and $\sigma_2$ can be found in \eqref{eqn:C_matrix}.

Similar to the CARP-MLN model, for the CARP-copula model, we also use a linear form of the covariates $\xvec_i$ to represent location parameters in marginal distributions as in \eqref{eqn:linear_tran}.
In the CARP-MLN model, we use a two dimensional vector $\muvec(\xvec_i) = [\mu_{1}(\xvec_{i}), \mu_{2}(\xvec_{i})]'$ to represent the location parameter of the lognormal distribution. While in the CARP-copula, $\mu_{1}(\xvec_{i})$ and $\mu_{2}(\xvec_{i})$ stand for the location parameters for the first and second marginal distributions, respectively.

\subsection{Properties of CARP}\label{subsection:properties.tprp}
For event gap time variable $\Wvec_i$, we define the survival function (sf), cdf and hazard function as follows. We denote the joint sf of $\Wvec_i$ as,
\begin{align}\label{eqn:bi.survival}
S(\vvec)=\pr(W_{i1}>v_1,  W_{i2}>v_2).
\end{align}
The joint cdf is $$F_{\Wvec}(\vvec)=\pr(W_{i1}\leq v_1,  W_{i2}\leq v_2),$$ and the corresponding joint probability density function (pdf) is denoted by $f_{\Wvec}(\vvec)$. According to \eqref{eqn:prp.renewal}, the joint pdf of the event gap time variable given all historical events $\Wvec_i|\F_{t_i}$ is given by
\begin{align}\label{eqn:cond.density}
f_{\Wvec}(\vvec|\avec_i, \xvec_i)=&\frac{f_{\Wvec}[a_1(c_1), a_2(c_2)]}{S(\avec_i)}, \quad v_j>a_j(t_i), \quad j=1, 2,
\end{align}
where $c_j=t_{l_{ij},j}+W_j, j=1,2$. The denominator in \eqref{eqn:cond.density} takes  age condition into account, while the numerator builds the relationship among the event gap time $W_j$, the event time $t_{l_{ij},j}$ and the age $a_j(c_j)$. Covariates $\xvec_i$ are used to adjust the gap time $\Wvec_i$ as discussed in Section \ref{sec:Net_dependence}.

For further discussions, let $\F_{t^{-}}$ be the event history up to time $t$. Note that for event type $j$, $T_{N_j(t^{-}),j}$ gives the most recent event time by time $t$, and the age variable prior to time $t$ is denoted as $A_{j}^{-}(t)=t-T_{N_j(t^{-}),j}$, which calculates the cumulative running time upon time $t$ since the last event. We use $\avec^{-}(t)=[a_{1}^{-}(t), a_{2}^{-}(t)]'$ to denote the age vector for the two event processes prior to time $t$. Hence, prior to an event time $t_i$, the age vector is denoted as $\avec_i^{-}=\avec^{-}(t_i)$. For example, at the initial time 0, the vector is $\avec_{0}^{-}=(0,0)'$. If the event type is $\delta_1=1$ at $t_1$, then $\avec_{1}^{-}=(t_1, t_1)'$. Note that $\avec_1$ updates the age to be zero for the corresponding event type at an event time, while $\avec_1^{-}$ does not. This notation is used to derive the likelihood in Section \ref{sec:par.est}, is also used to define the hazard function below.

In the literature, the sub-intensity function (i.e., cause-specific event intensity function) is often used to characterize an event process. In particular, the sub-intensity function for event type $j$ is defined as
\begin{align}\label{eqn:sub.intensity.def}
h_j(t)=\lim_{\Delta t\to0}\ffrac{\pr[T\in (t, t+\Delta t), \Delta=\delta_{j}|\F_{t^{-}}]}{\Delta t},
\end{align}
where $T$ is the event time, and $\Delta$ is the event type. The sub-cumulative intensity function is $H_j(t)=\int_{0}^th_j(s)ds$. The hazard function and cumulative hazard function for the system are calculated as the sum of corresponding functions for the two event types, $$h(t)=\sum_{j=1}^2h_j(t) \quad \textrm{and} \quad H(t)=\sum_{j=1}^2H_j(t).$$
The sub-intensity function in \eqref{eqn:sub.intensity.def} is calculated as
\begin{align}\label{eqn:hkt}
h_j(t)=\frac{D_j[\avec^{-}(t)]}{S[\avec^{-}(t)]},\quad
\textrm{ where }\quad
D_j(\widetilde{\vvec})=-\frac{\partial S(\vvec)}{\partial v_{j}}\bigg|_{\vvec=\widetilde{\vvec}},
\end{align}
and $\widetilde{\vvec}=(\widetilde{v}_1, \widetilde{v}_2)'$ is a vector with two components. More detail about the calculation of $D_j(\widetilde{\vvec})$ under different models will be discussed in Section \ref{sec:par.est}.


\section{Parameter Estimation}\label{sec:par.est}

The maximum likelihood (ML) approach is used to estimate model parameters. Parameters in the model include parameters in the joint distribution function, the copula function and the linear covariate transformation function. Given all the event history $\F_{\tau}$, the likelihood function is constructed as follows:
\begin{align}\label{eqn:likelihood}
L(\thetavec |\F_{\tau})=\prod_{i=1}^{n}L_{i}(\thetavec | \avec_i, \xvec_i),
\end{align}
where

\begin{align}\label{eqn:Li}
L_{i}(\thetavec | \avec_i, \xvec_i)=\pr[T_{i}\in(t_{i}, t_{i}+\Delta t), \Delta_{i}=\delta_{i}|\F_{t_{i-1}}], \quad \textrm{for} \quad i=1,\cdots, n.
\end{align}
Let $\thetavec $ denote the vector of all the parameters in the model.  The estimated parameters $\thetavechat$ are asymptotically normally distributed based on the large sample ML theory \citep{casella2002statistical}. The calculation of the likelihood $L_{i}$ for the proposed CARP models is shown below.

For any observed recurrent process with $n$ total events, the likelihood contribution for $i=1,\cdots, n$ in \eqref{eqn:Li} is given by
\begin{align}\label{eqn:Li2}
L_{i}(\thetavec | \avec_i, \xvec_i) =\frac{D_{\delta_i}(\avec_i^{-})}{S(\avec_{i-1})}.
\end{align}
In \eqref{eqn:Li2}, the quantity $D_{\delta_i}(\avec_i^{-})$, which is introduced in \eqref{eqn:hkt}, is the partial derivative of the bivariate sf $S(\avec_i^{-})$. The covariates $\xvec_i$ are used to adjust the distribution of $\Wvec_i$. Likelihood calculations so far are the same for both the CARP-MLN and CARP-copula models. However, we need different ways to calculate $D_j(\avec_i^{-})$ for the two CARP models based on how the survival functions are calculated, which are detailed below.

For the CARP-MLN model, the sf is calculated in a closed form as discussed in \eqref{eqn:bi.survival}, and the partial derivative with regard to the $j$th event type can be written as
\begin{align}\label{eqn:partial.derivative}
D_j(\avec_i^{-})=-\frac{\partial S(v_{j}, v_{j'})}{\partial v_{j}}\bigg|_{\vvec=\avec_i^{-}} = f_{j}[a^{-}_{j}(t_{i})] \times \Pr[W_{j'} \ge a^{-}_{j'}(t_{i}); j' \neq j | W_{j} = a^{-}_{j}(t_{i})],
\end{align}
where $f_{j}[a^{-}_{j}(t_{i})]$ is the $j$th marginal density function from a bivariate lognormal distribution. In \eqref{eqn:partial.derivative}, the conditional probability can be calculated from the conditional normal distribution with a logarithm transformation. Calculation details can be found in Appendix~\ref{Conditional Lognormal Probability}.

For the CARP-copula model, the likelihood function is calculated based on the relationship between the bivariate sf $S(\avec_i^{-})$ and the cdf $F(\avec_i^{-})$. In bivariate cases, the sf and cdf have the following relationship:
$$S[a_{1}(t_{i}), a_{2}(t_{i})] = 1 - F[a_{1}(t_{i}), \infty] - F[\infty, a_{2}(t_{i})] + F[a_{1}(t_{i}), a_{2}(t_{i})],$$
where the joint cdf can be calculated by the copula as in \eqref{eqn:sklar}. The calculated likelihood will vary for different choices of the marginal distribution and the copula function. In our case, we chose the Gumbel copula as described in Section~\ref{sec:Net_dependence}. The ML estimates $\thetavechat$ are obtained by maximizing the likelihood function in \eqref{eqn:likelihood}.

\section{Simulation Study}\label{sec:simulation}

This section describes the simulation study we conducted for evaluating the proposed method. We simulated bivariate recurrent system data based on different models and parameter values. The performance of the proposed method was then evaluated based on the simulated data. Different parameter values were considered in the simulation for both the CARP-MLN and the CARP-copula models to understand the impact of the model parameters. We evaluated the model goodness of fit by calculating the average AIC and the mean squared error (MSE) of the estimated model parameters. The goal is to demonstrate the performance of the proposed method and the improvement by using the covariate adjustment.

\subsection{Simulation Setting}

We used both the CARP-MLN and CARP-copula models to simulate the data. For each generated data set, both models were fitted to the data and results were evaluated. The model used to generate data is referred to as the true model, while the models used to fit the data are referred to as the fitted models. We explored changing the sample sizes ($n$), the Kendall's tau value, the scale parameters ($\sigma_j$, $j=1, 2$), and the linear transformation matrix ($\Bvec$) to understand their impacts on the analysis.

First, to understand the impact of sample size, we varied the sample size at $n=$200, 500, 1000 and 2000 when generating the data using each model. For the CARP-copula model, the lognormal marginal distributions were used for both event types along with the Gumbel copula function. The location and scale parameters for lognormal marginal distributions in the true model were specified at $(\mu_{1} = 1, \sigma_{1} = 0.25)$ and $(\mu_{2} = 1.5, \sigma_{2} = 0.25)$. For the linear transformation matrix $\Bvec$, we used the 2 $\times$ 2 matrix below
\begin{align}\label{eqn:B_matrix}
\Bvec = \left(\begin{array}{cc} 1.5 & 0 \\ 0 & 0.1 \end{array}\right).
\end{align}
We chose to set the Gumbel copula parameter $\alpha$ at 1.5. When generating the data based on the CARP-MLN model, we used the same location and scale parameters for the marginal lognormal distributions, and adjusted the correlation parameter $\eta$ to obtain the same Kendall's tau as in the CARP-copula model.

Second, to evaluate the effect of the Gumbel copula coefficient $\alpha$, we varied its value at 1, 1.12, 1.5 and 2.22 by changing the Kendall's tau parameter in the true models, while keeping all other parameters the same with the sample size fixed at $n=1000$. Considering that the CARP-MLN and CARP-copula models quantify the Kendall's tau differently, we used $\eta$ at 0, 0.0443, 0.1445 and 0.299 for the CARP-MLN model, so that the Kendall's tau from both CARPs are varied at 0, 0.11, 0.33 and 0.55.

We also explored the impact of the scale parameters by letting $\sigma_{1}$ and $\sigma_{2}$
vary at 0.35, 0.3, 0.25 to 0.20 for the marginal lognormal distributions of the CARP-Copula model while keeping sample size at $n=1000$ and the Kendall's tau was kept at 0.33. Similarly for the CARP-MLN model, covariance matrices were adjusted to align with the marginal distributions used in the CARP-Copula model.

Lastly, the effect of covariate adjustment was studied by using different types of $\Bvec$. True models used both non-zero and zero $\Bvec$ matrices. In the non-zero $\Bvec$ case, we used the coefficient matrix in \eqref{eqn:B_matrix} to generate data. In the zero $\Bvec$ case, $\Bvec = \boldsymbol{0}$ was used. The Kendall's tau was set to be 0.33, while other parameters are the same as ones in the sample size case. A sample size of 1000 is used across true models.

\subsection{Simulation Results}

The simulation results under different true models and fitted models are summarized in this section. For each true model, the model performance was summarized over 1000 simulated data sets. The simulation size was chosen to ensure a reliable analysis result while balancing the computing time needed to evaluate a broad number of scenarios. Further increasing the simulation size would not result in a meaningful change in the evaluated summary statistics and the drawn conclusions. Under each fitted model, the average AIC was computed. In order to evaluate the performance of parameter estimates, we calculated the MSE of the location parameter $\mu$, scale parameter $\sigma$ and the linear transformation parameter $\Bvec$ for different sample sizes.

\begin{table}
	\caption{Average AIC from CARP-MLN and CARP-copula calculated by 1000 repeated samples on true models generated by both CARP-MLN and copula models. Sample sizes from true models are changed from  200, 500, 1000 and 2000.}
	\centering
	\begin{tabular}{P{2cm}|P{2cm}|P{2cm}P{2cm}|P{2cm}P{2cm}}
		\hline
		\hline
		\multicolumn{2}{c|}{True Model} & \multicolumn{2}{c|}{Copula Generation} &  \multicolumn{2}{c}{MLN Generation} \\
		\hline
		\multicolumn{2}{c|}{Fitted Model} & MLN & Copula & MLN & Copula\\
		\hline
		\multirow{4}{2cm}{\centering Sample size $n$}  & 200 & 702.0 & 722.5 & 722.9 & 721.9\\
		
		& 500& 1755.4 & 1751.3 & 1806.1 & 1808.7 \\
		
		&  1000 & 3513.7 & 3504.9  & 3615.4 & 3622.3 \\
		
		& 2000 & 7027.7  &  7014.3 & 7220.0 & 7234.8 \\
		
		\hline
		\hline
	\end{tabular}
	
	\label{table:result1}
\end{table}

\begin{table}
	\caption{Average AIC by CARP-MLN and copula under different Kendall's tau. $\alpha$ and $\eta$ are used to adjust the Kendall's tau in the true models for CARP-MLN and CARP-copula, respectively.}
	\centering
	\begin{tabular}{P{2cm}|P{2cm}|P{2cm}P{2cm}|P{2cm}P{2cm}}
		\hline
		\hline
		\multicolumn{2}{c|}{True Model}  & \multicolumn{2}{c|}{Copula Generation} &  \multicolumn{2}{c}{MLN Generation} \\
		\hline
		\multicolumn{2}{c|}{Fitted Model} & MLN & Copula & MLN & Copula\\
		
		\hline
		\multirow{4}{2cm}{\centering Kendall's tau ($\tau$)}  & 0& 3472.2 & 3459.5 & 3459.4 & 3455.7 \\
		
		& 0.11 & 3510.5  & 3506.9 & 3489.3 & 3490.8 \\
		
		&0.33 & 3513.7 & 3504.9 & 3611.7 & 3618.6 \\
		
		&0.55 & 3445.6 & 3439.7 & 3837.3 & 3852.3 \\
		
		\hline
		\hline
	\end{tabular}
	
	\label{table:result2}
\end{table}

\begin{table}
	\caption{The Average AIC by CARP-MLN and copula under different scale parameters. $\sigma_1$ and $\sigma_2$ are set equal in the true models for CARP-MLN and CARP-copula.}
	\centering
	\begin{tabular}{P{2cm}|P{2cm}|P{2cm}P{2cm}|P{2cm}P{2cm}}
		\hline
		\hline
		\multicolumn{2}{c|}{True Model}  & \multicolumn{2}{c|}{Copula Generation} &  \multicolumn{2}{c}{MLN Generation} \\
		\hline
		\multicolumn{2}{c|}{Fitted Model} & MLN & Copula & MLN & Copula\\
		
		\hline
		\multirow{4}{2cm}{\centering Scale parameter ($\sigma_1$ and $\sigma_2$)}  & 0.35& 4084.7 & 4071.7 & 4148.9& 4158.1 \\
		
		& 0.30 & 3747.4  & 3736.1 & 3820.0 & 3826.7 \\
		
		&0.25 & 3513.7 & 3504.9 & 3615.4 & 3622.3 \\
		
		&0.20 & 2891.6 & 2882.7 & 2980.9 & 2983.3 \\
		
		\hline
		\hline
	\end{tabular}
	
	\label{table:result5}
\end{table}

\begin{table}
	\caption{Average AIC from different true models and fitted models to evaluate effect of covariate adjustment. True and fitted models are copula and MLN with or without coefficient $\Bvec$.}
	\label{Table:misspecfication}
	\centering
	\begin{tabular}{ P{2.1cm} | P{2.2cm} | P{2.2cm} P{2.1cm} | P{2.2cm} P{2.1cm}}
		\hline
		\hline
		\multicolumn{2}{c|}{ \multirow{2}{*}{\centering \backslashbox{True Model}{Fitted Model}} }  & \multicolumn{2}{c|}{Copula} & \multicolumn{2}{c}{MLN}\\
		\hhline{~~~~~~}
		\multicolumn{2}{c|}{}  & Non-zero $\Bvec$ &  Zero $\Bvec$  & Non-zero $\Bvec$ & Zero $\Bvec$ \\

		\hline
		\multirow{2}{*}{ Copula} & Non-zero $\Bvec$ & 3505.6 & 4069.9 & 3514.4 & 4064.8 \\
		\hhline{~~~~~~}
		& Zero $\Bvec$ & 2480.5 & 2485.1 & 2492.9 & 2489.0\\
		\hline
		\multirow{2}{*}{ MLN} & Non-zero $\Bvec$ & 3617.7 & 4186.2 & 3610.8 & 4180.0\\
		\hhline{~~~~~~}
		& Zero $\Bvec$ & 2572.9 & 2568.7 & 2565.3 & 2561.1\\
		\hline
		\hline
	\end{tabular}

	\label{table:result4}
\end{table}

\begin{figure}
	\centering
	\begin{subfigure}[h]{0.5\textwidth}
		\centering
		\includegraphics[width=\textwidth, height=6cm]{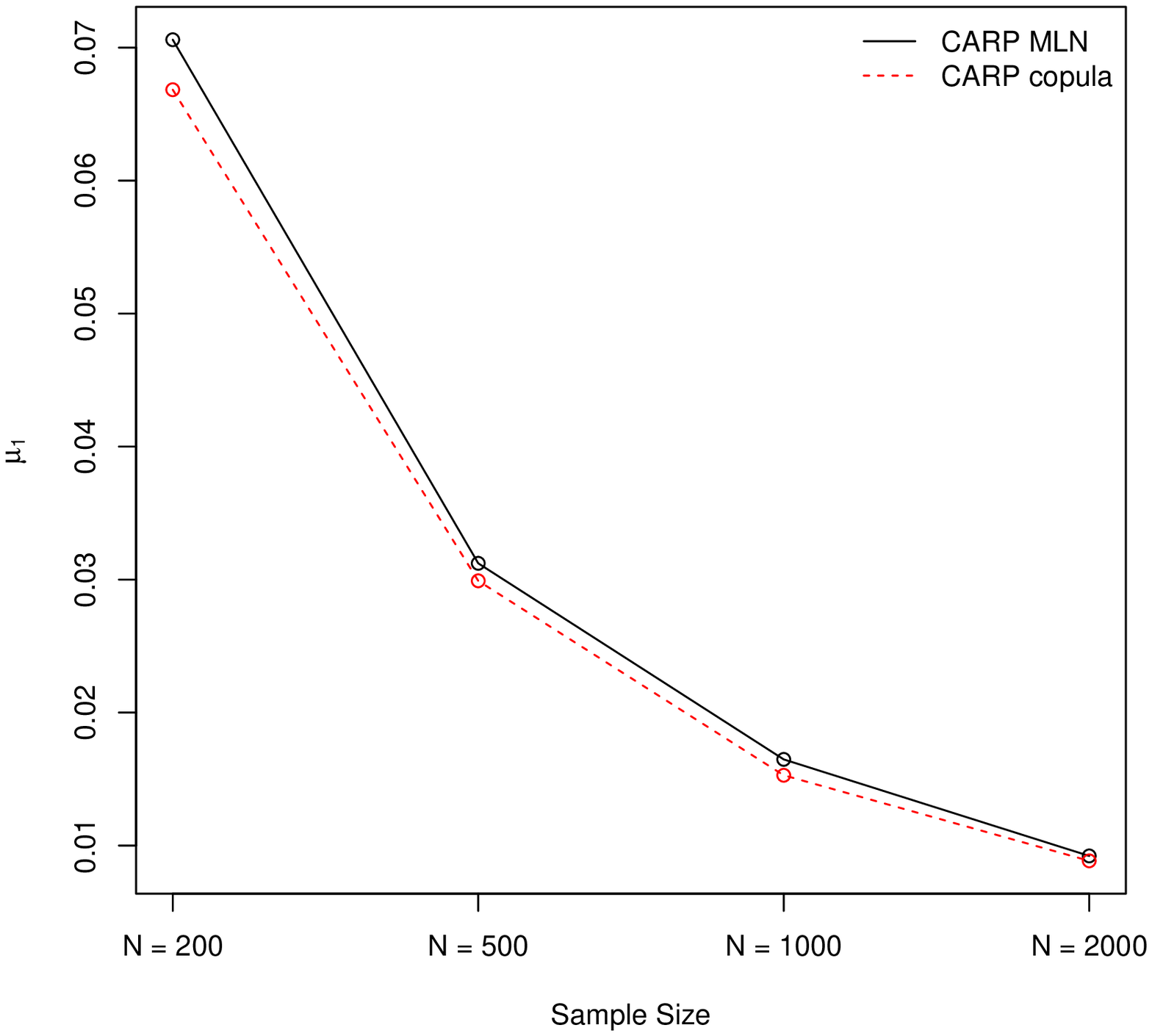}
	\end{subfigure}%
	~
	\begin{subfigure}[h]{0.5\textwidth}
		\centering
		\includegraphics[width=\textwidth, height=5.9cm]{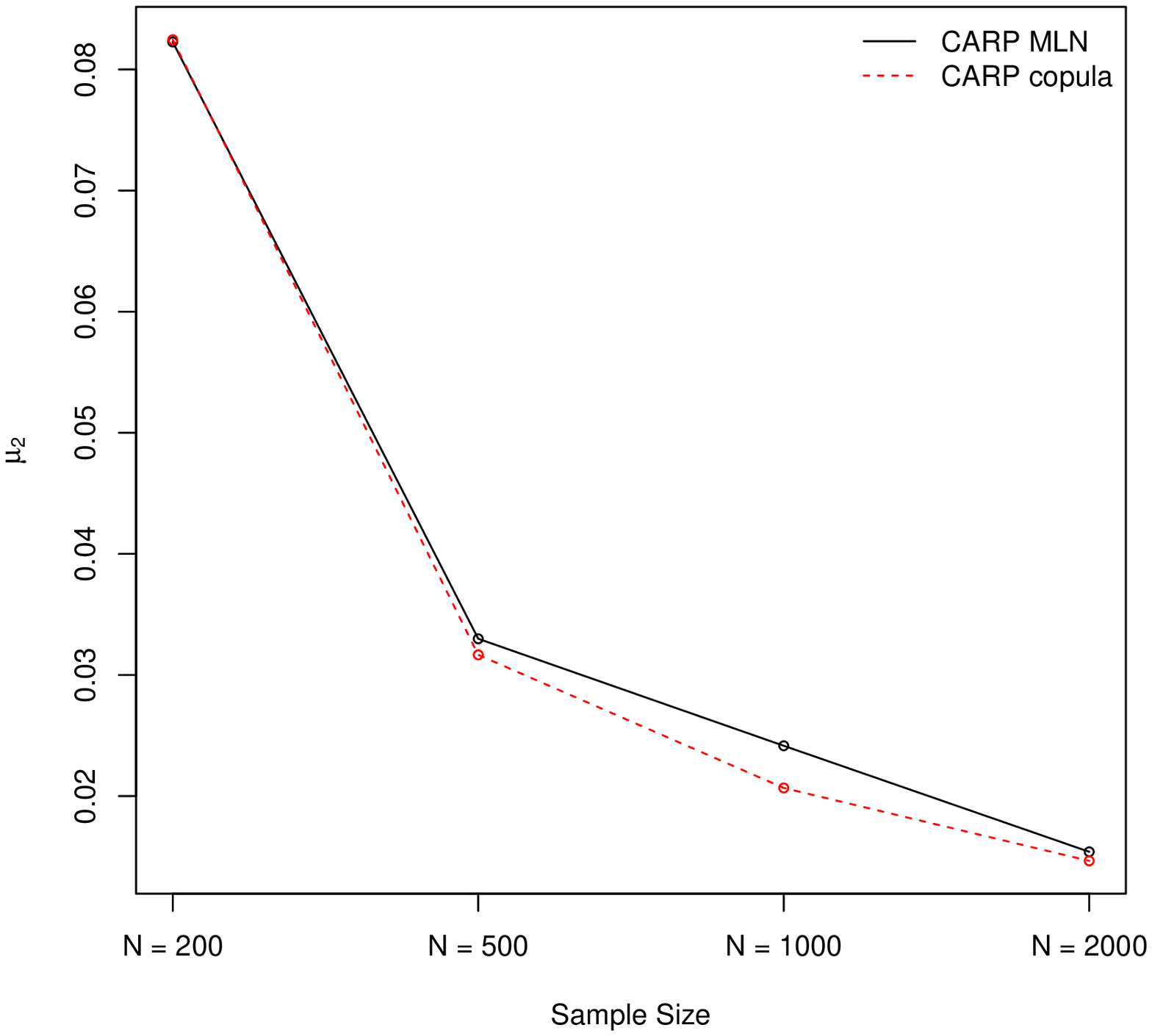}
	\end{subfigure}%
	
	~\\[-4ex]
	\begin{subfigure}[h]{0.5\textwidth}
		\centering
		\includegraphics[width=\textwidth, height=5.9cm]{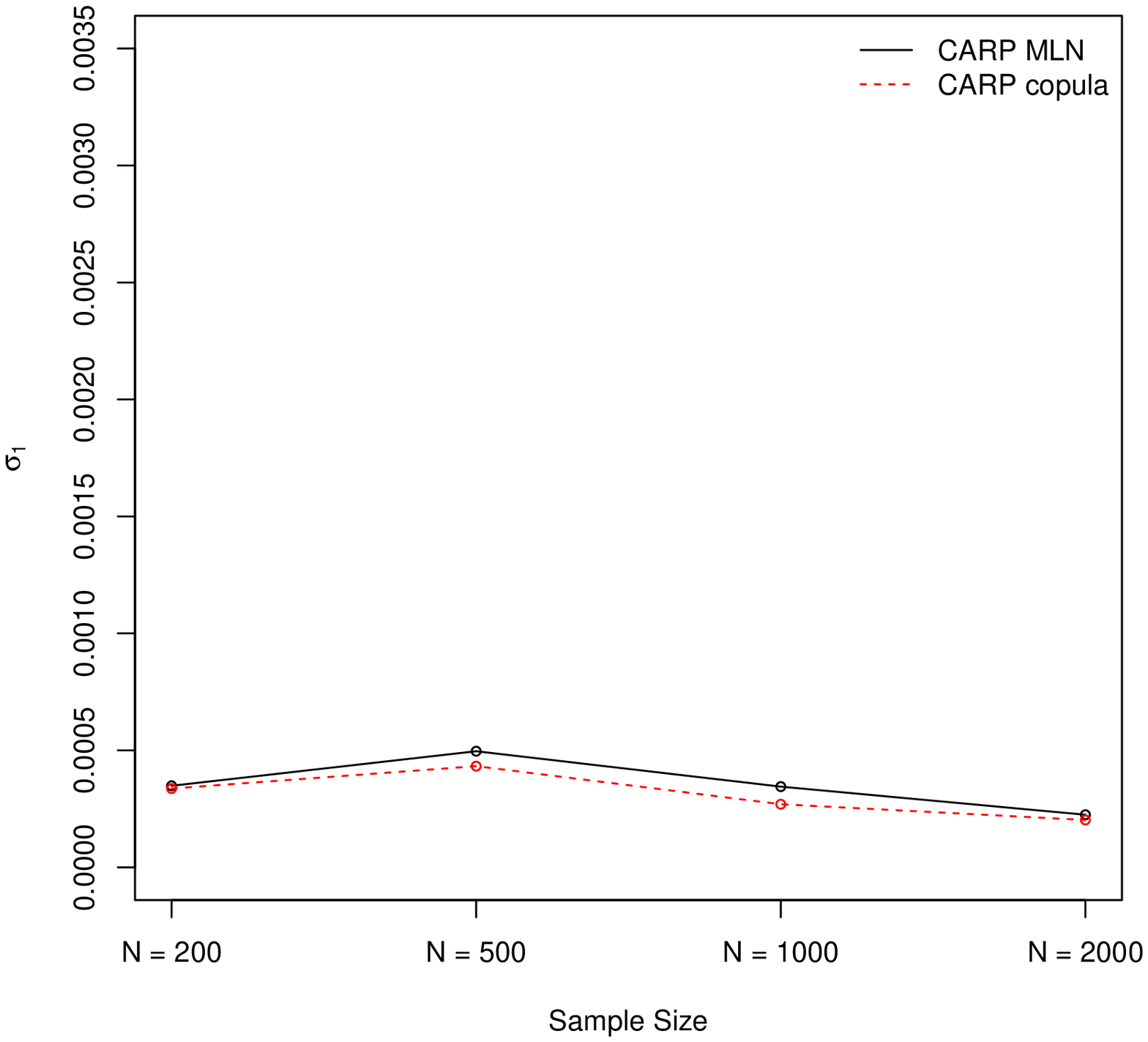}
	\end{subfigure}%
	~
	\begin{subfigure}[h]{0.5\textwidth}
		\centering
		\includegraphics[width=\textwidth, height=5.9cm]{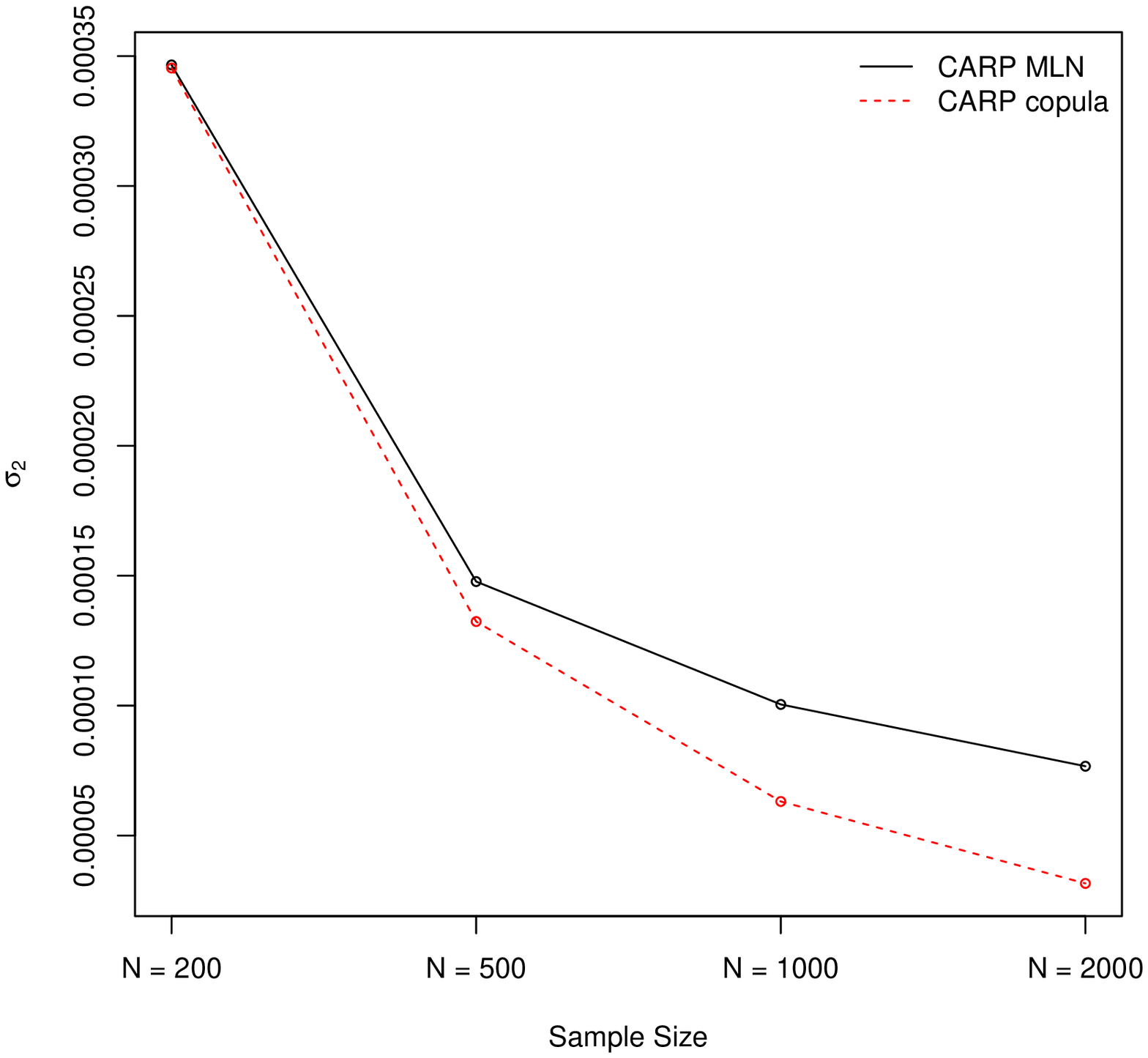}
		
	\end{subfigure}%
	
	~\\[-4ex]
	\begin{subfigure}[h]{0.5\textwidth}
		\centering
		\includegraphics[width=\textwidth, height=5.9cm]{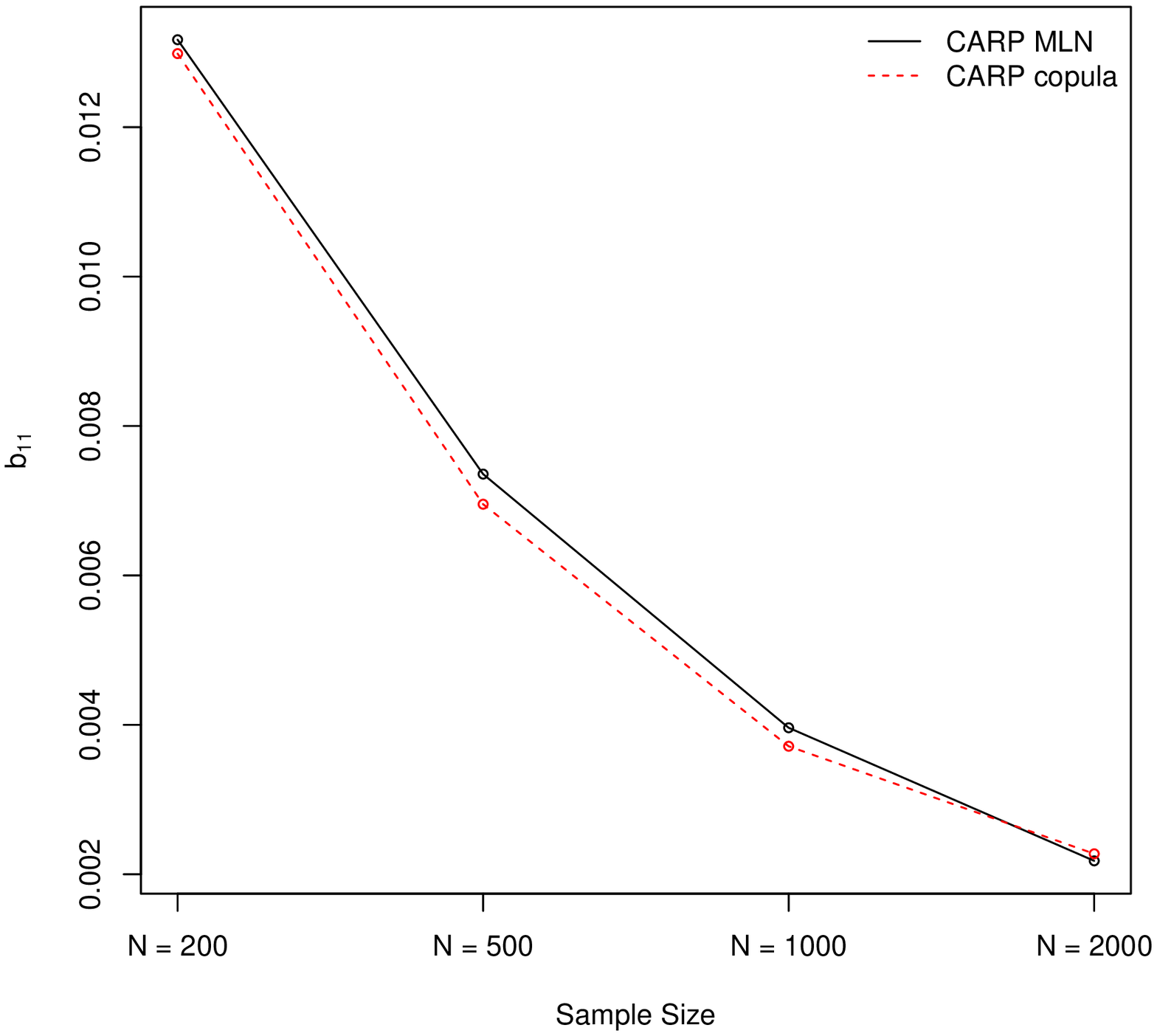}
		
	\end{subfigure}%
	~
	\begin{subfigure}[h]{0.5\textwidth}
		\centering
		\includegraphics[width=\textwidth, height=5.9cm]{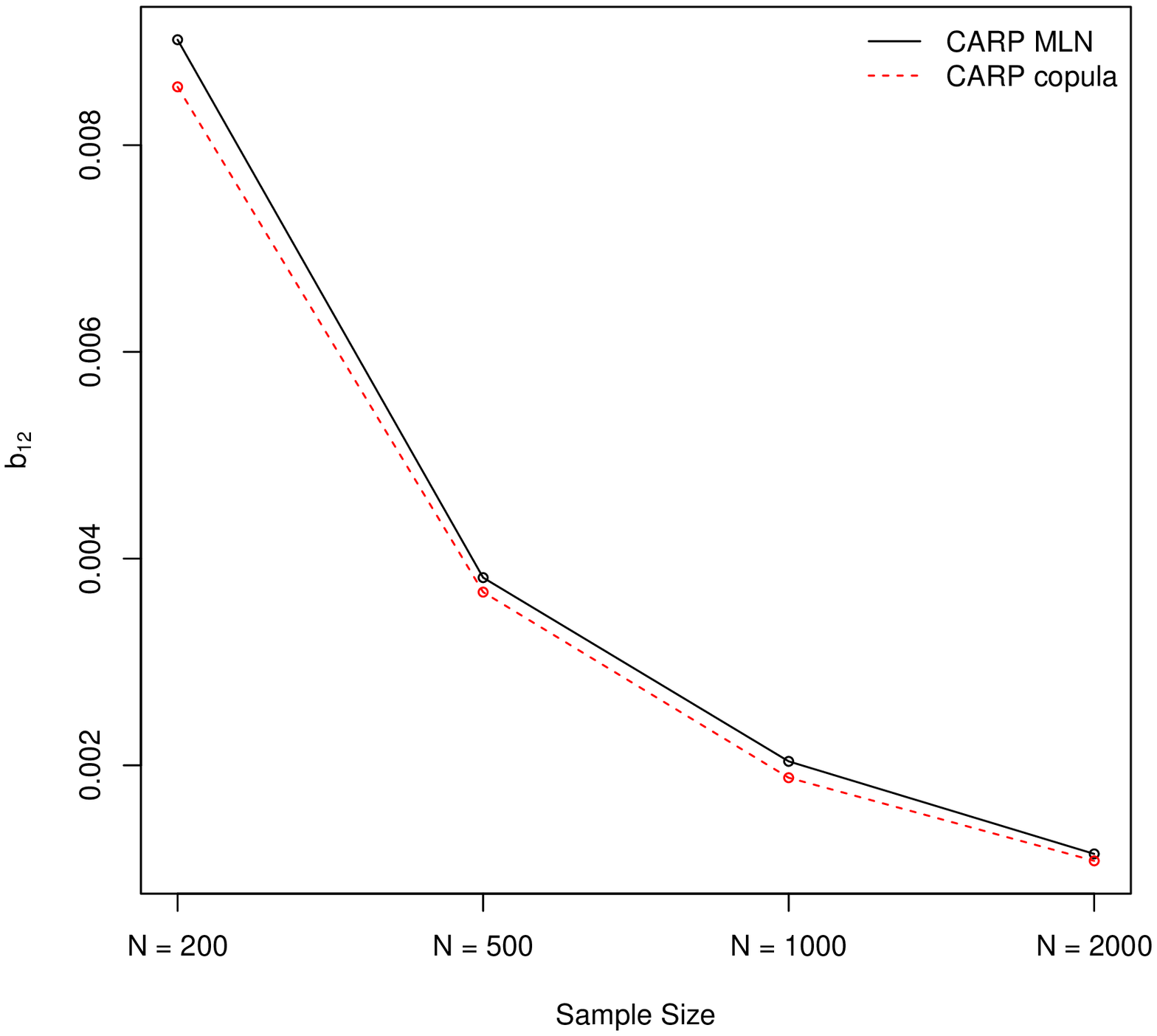}
		
	\end{subfigure}%
	~\\[-4ex]
	\begin{subfigure}[h]{0.5\textwidth}
		\centering
		\includegraphics[width=\textwidth, height=5.9cm]{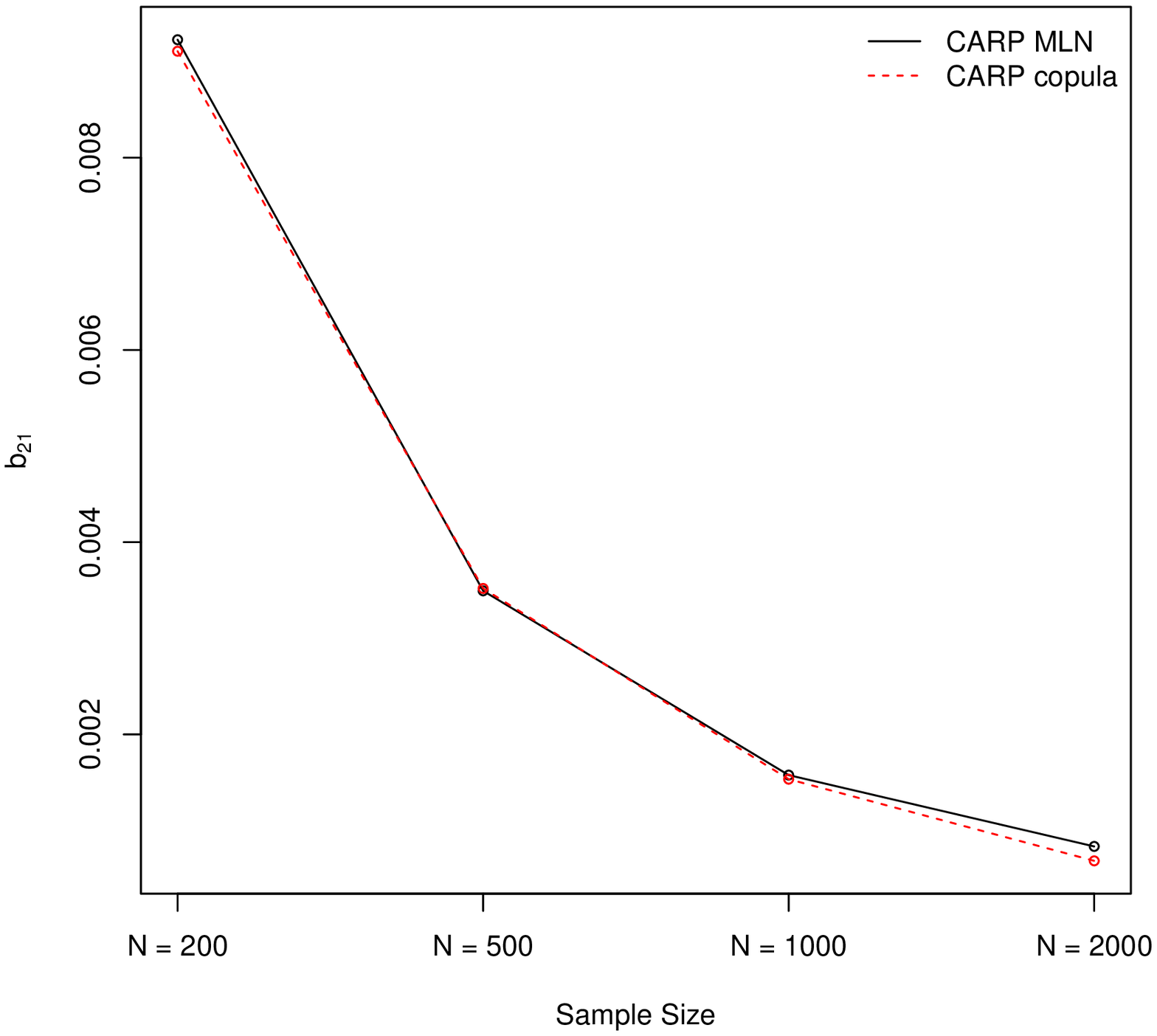}
		
	\end{subfigure}%
	~
	\begin{subfigure}[h]{0.5\textwidth}
		\centering
		\includegraphics[width=\textwidth, height=5.9cm]{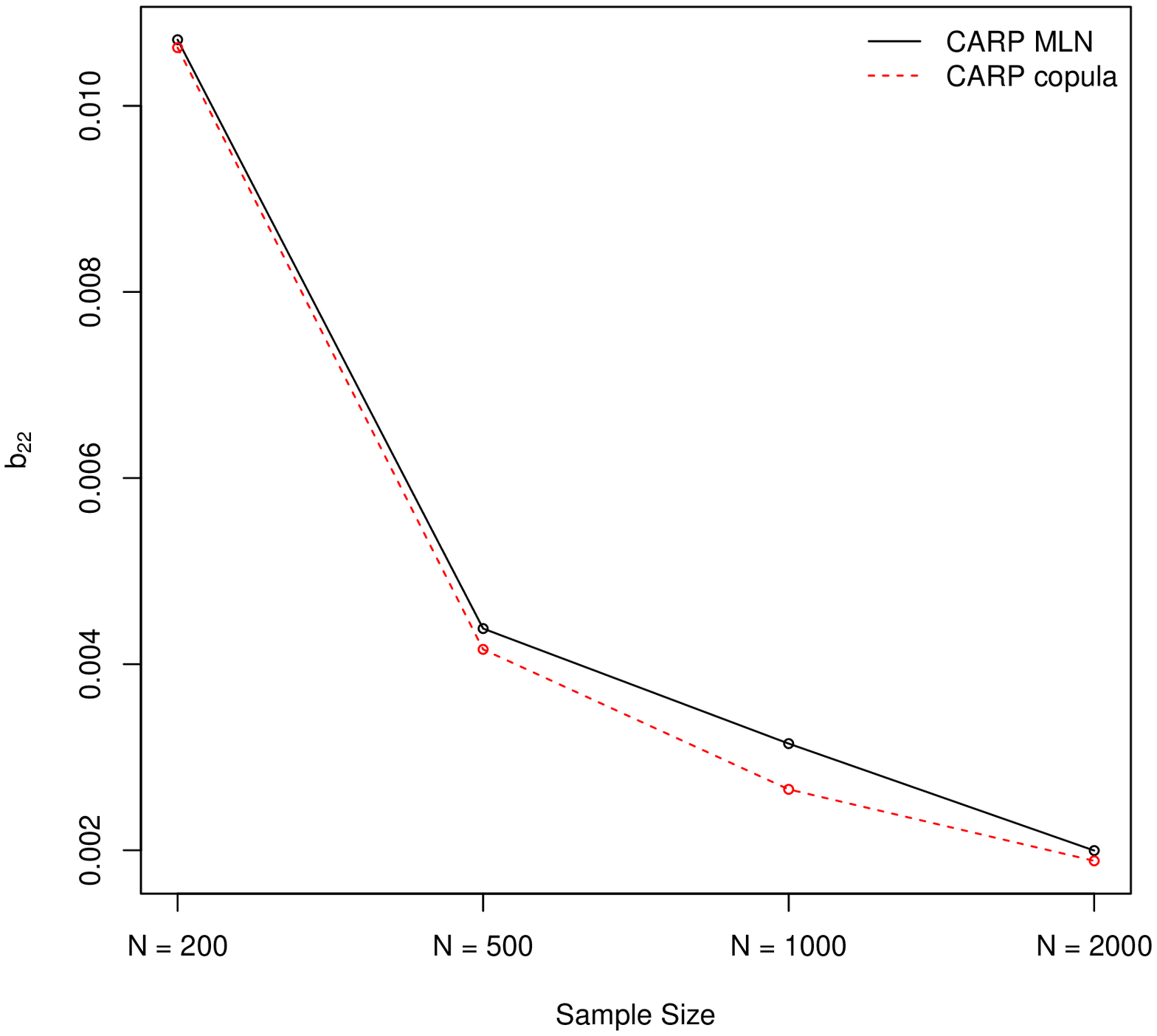}
		
	\end{subfigure}%
	~
	\caption{MSE for the location and parameter $\mu_1$, $\mu_2$, scale parameter $\sigma_1$, $\sigma_2$ and coefficient $\Bvec$, calculated by both CARP-MLN and copula model with different sample sizes. The true model is CARP-copula with lognormal marginal distributions.}
	\label{Fig:result1}
\end{figure}

\begin{figure}
	\centering
	\begin{subfigure}[h]{0.5\textwidth}
		\centering
		\includegraphics[width=\textwidth, height=5.9cm]{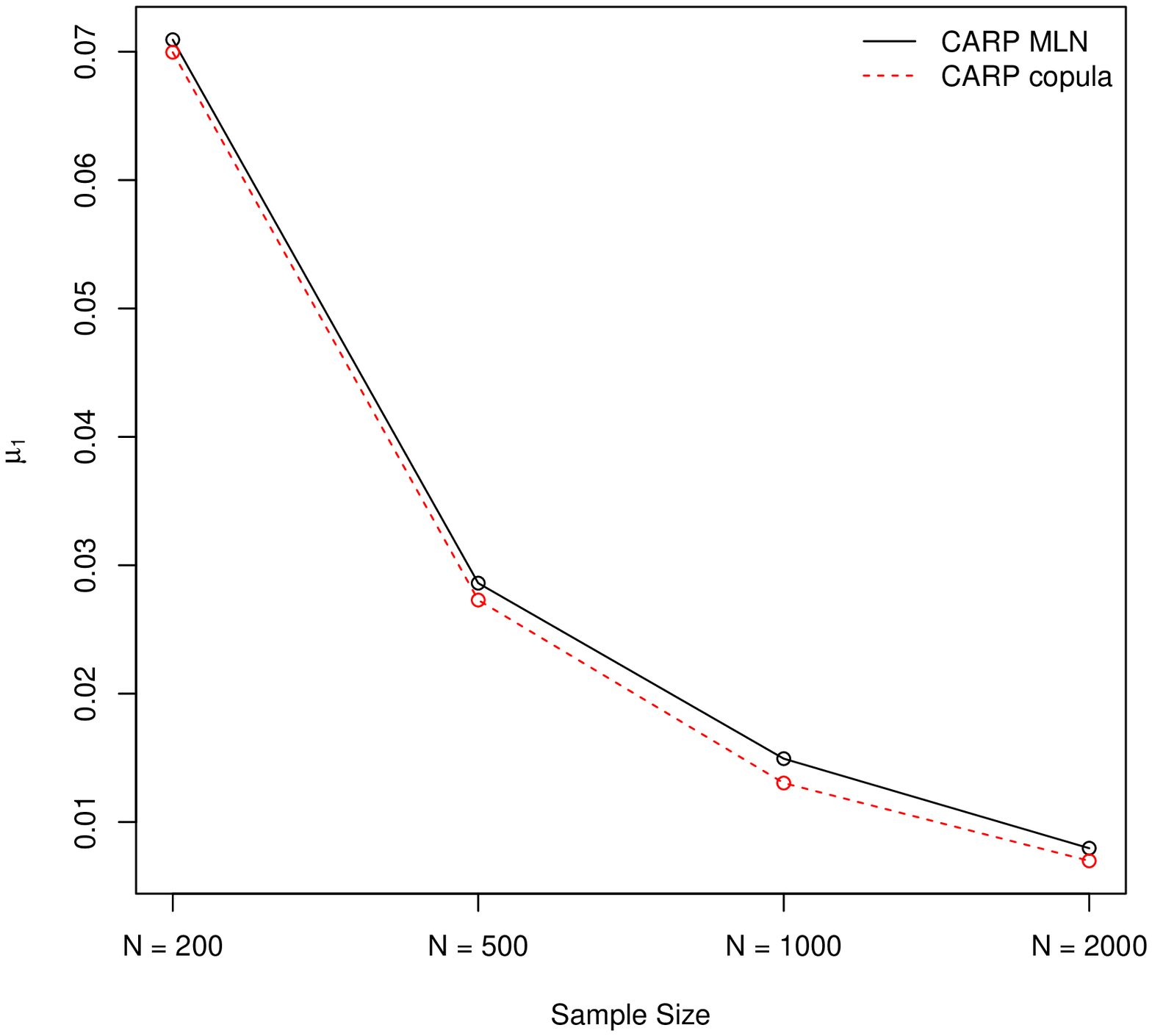}
	\end{subfigure}%
	~
	\begin{subfigure}[h]{0.5\textwidth}
		\centering
		\includegraphics[width=\textwidth, height=5.9cm]{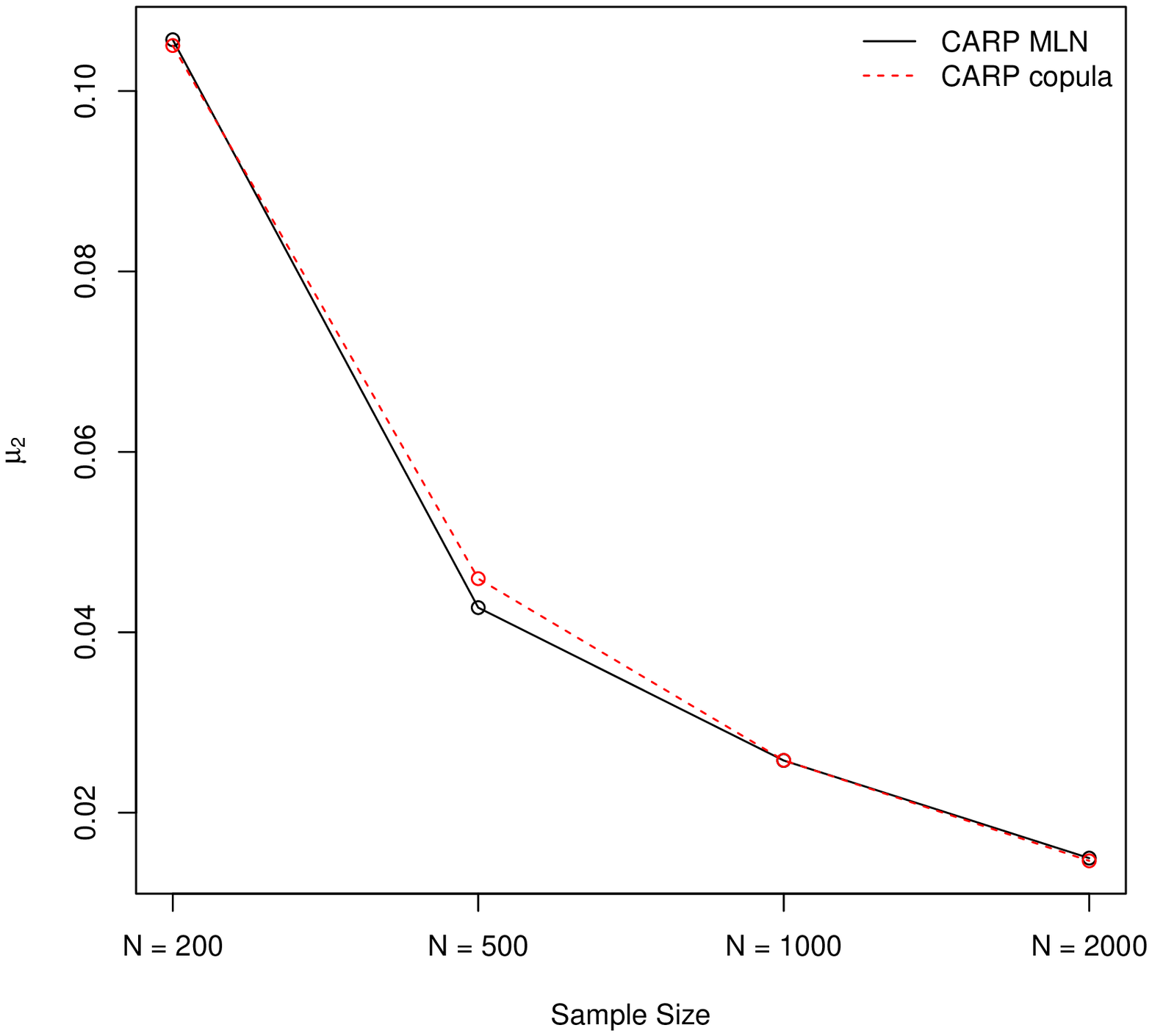}
	\end{subfigure}%
	
	~\\[-4ex]
	\begin{subfigure}[h]{0.5\textwidth}
		\centering
		\includegraphics[width=\textwidth, height=5.9cm]{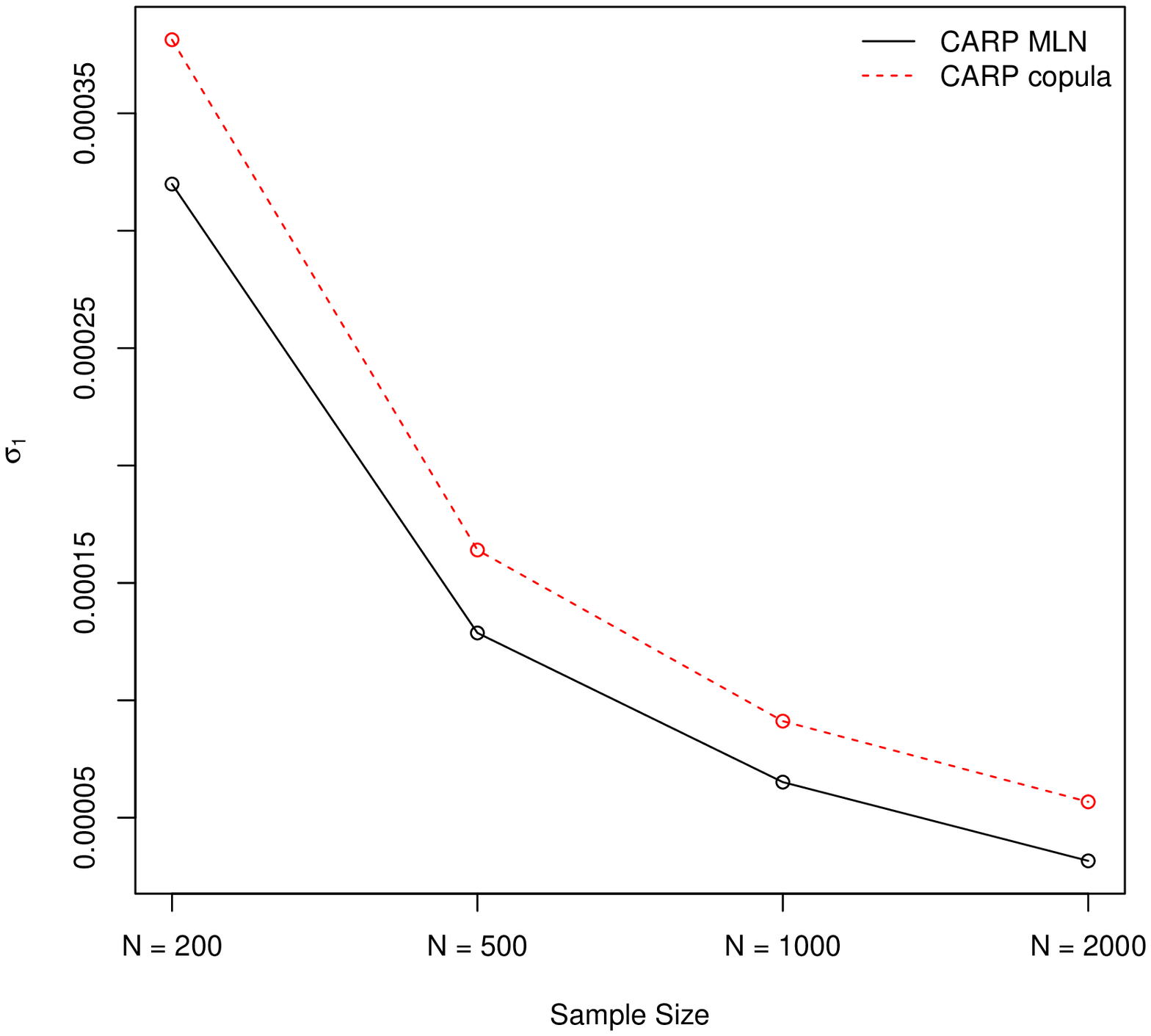}
	\end{subfigure}%
	~
	\begin{subfigure}[h]{0.5\textwidth}
		\centering
		\includegraphics[width=\textwidth, height=5.9cm]{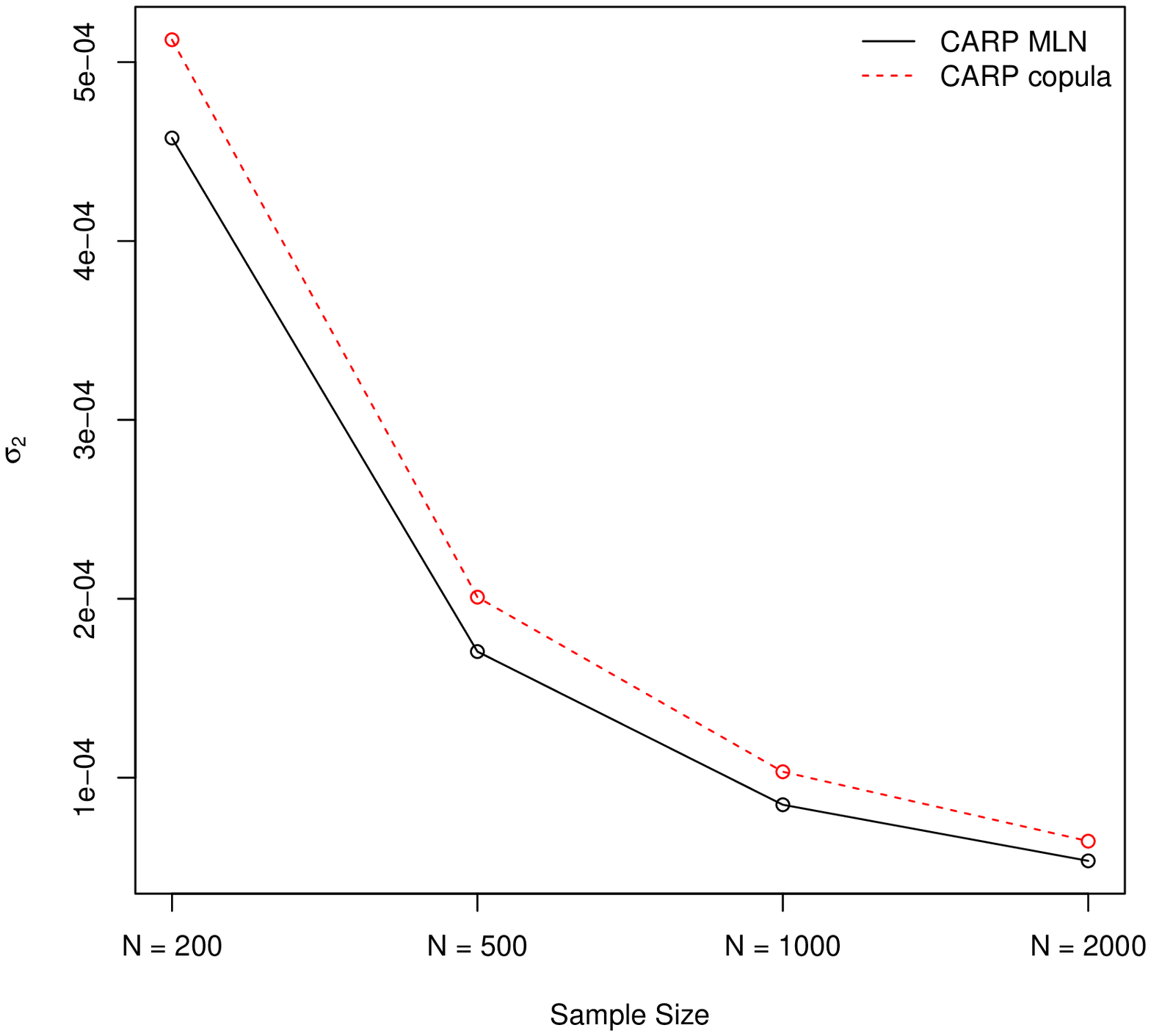}
		
	\end{subfigure}%
	
	~\\[-4ex]
	\begin{subfigure}[h]{0.5\textwidth}
		\centering
		\includegraphics[width=\textwidth, height=5.9cm]{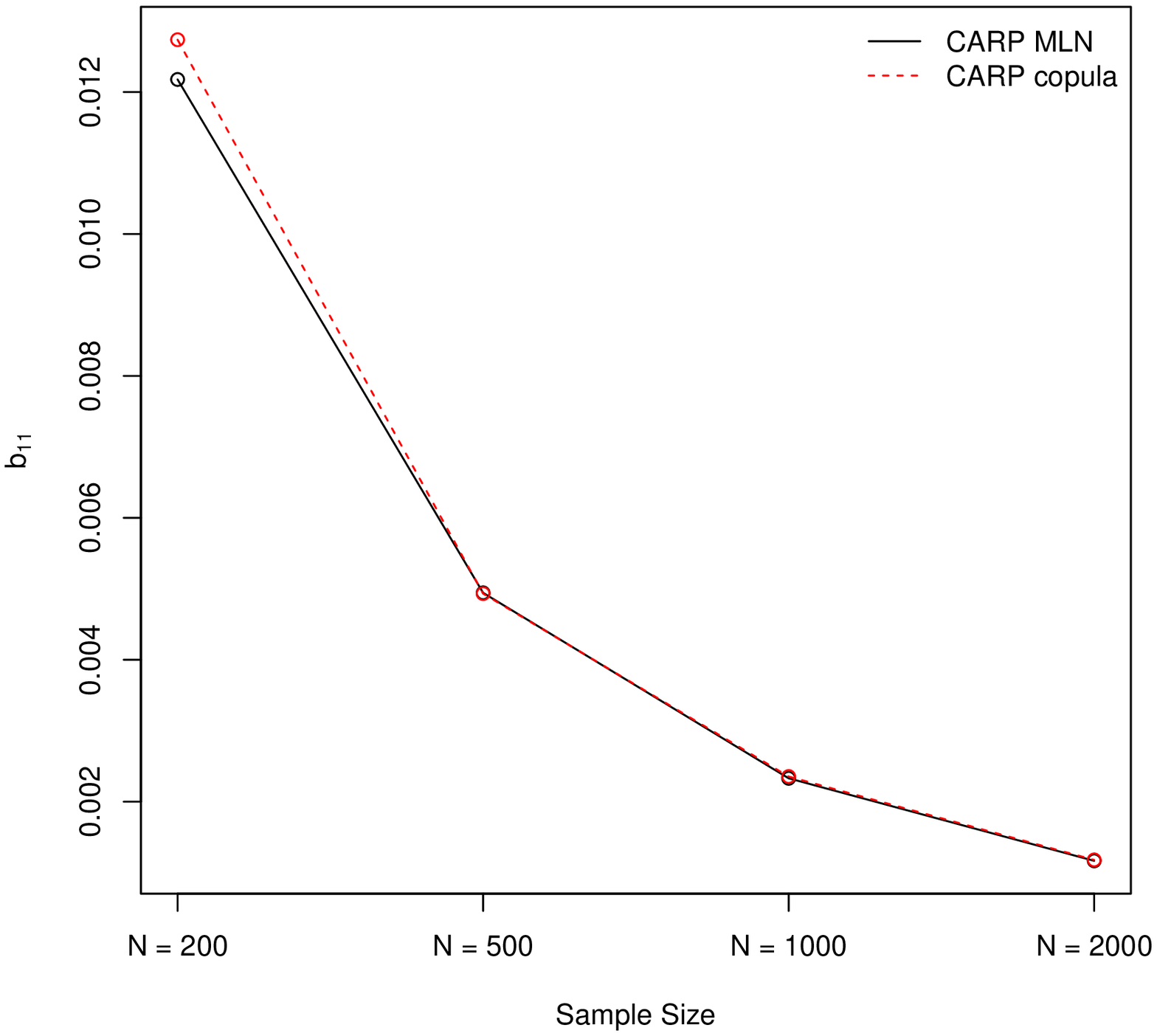}
		
	\end{subfigure}%
	~
	\begin{subfigure}[h]{0.5\textwidth}
		\centering
		\includegraphics[width=\textwidth, height=5.9cm]{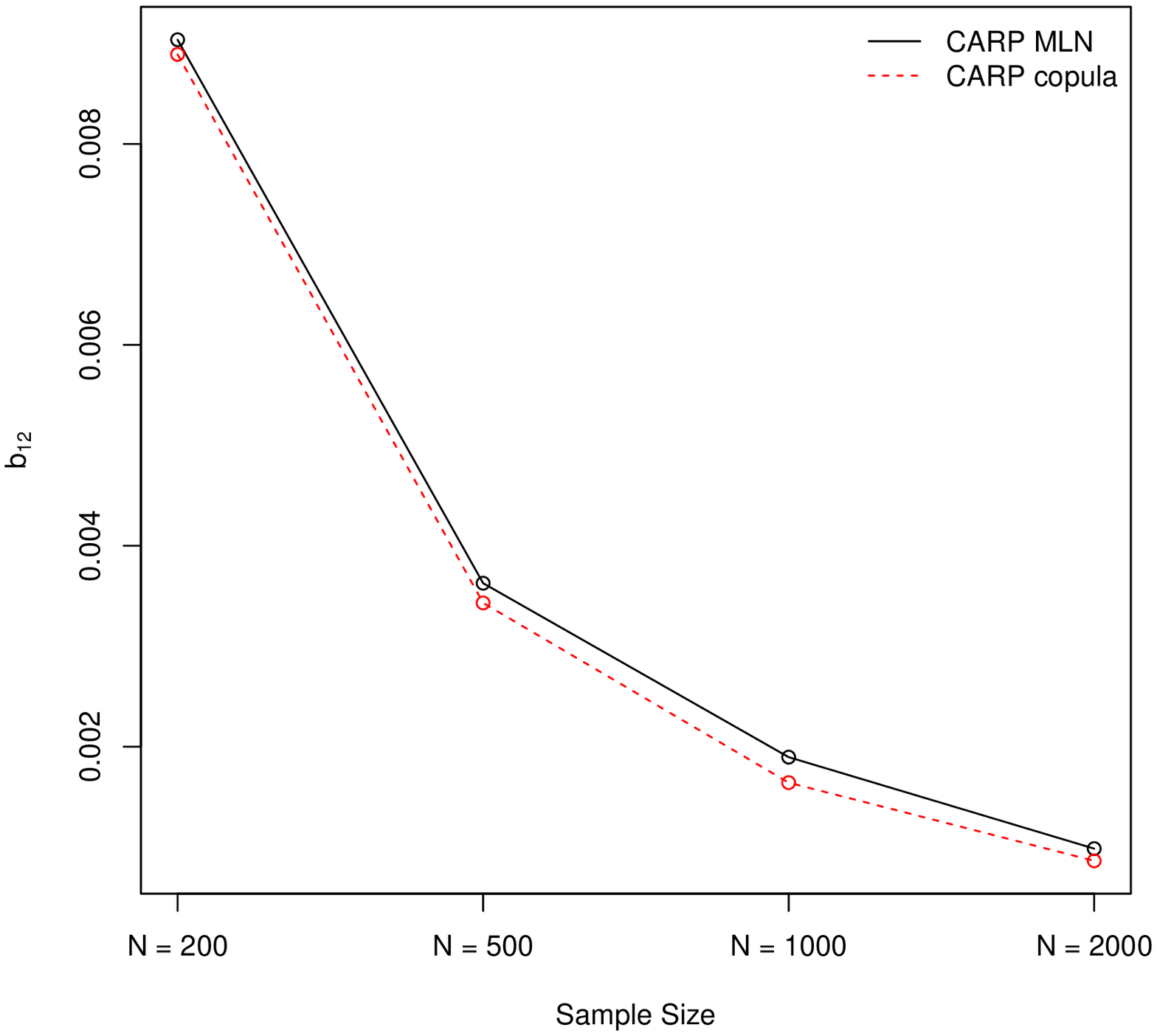}
		
	\end{subfigure}%
	
	~\\[-4ex]
	\begin{subfigure}[h]{0.5\textwidth}
		\centering
		\includegraphics[width=\textwidth, height=5.9cm]{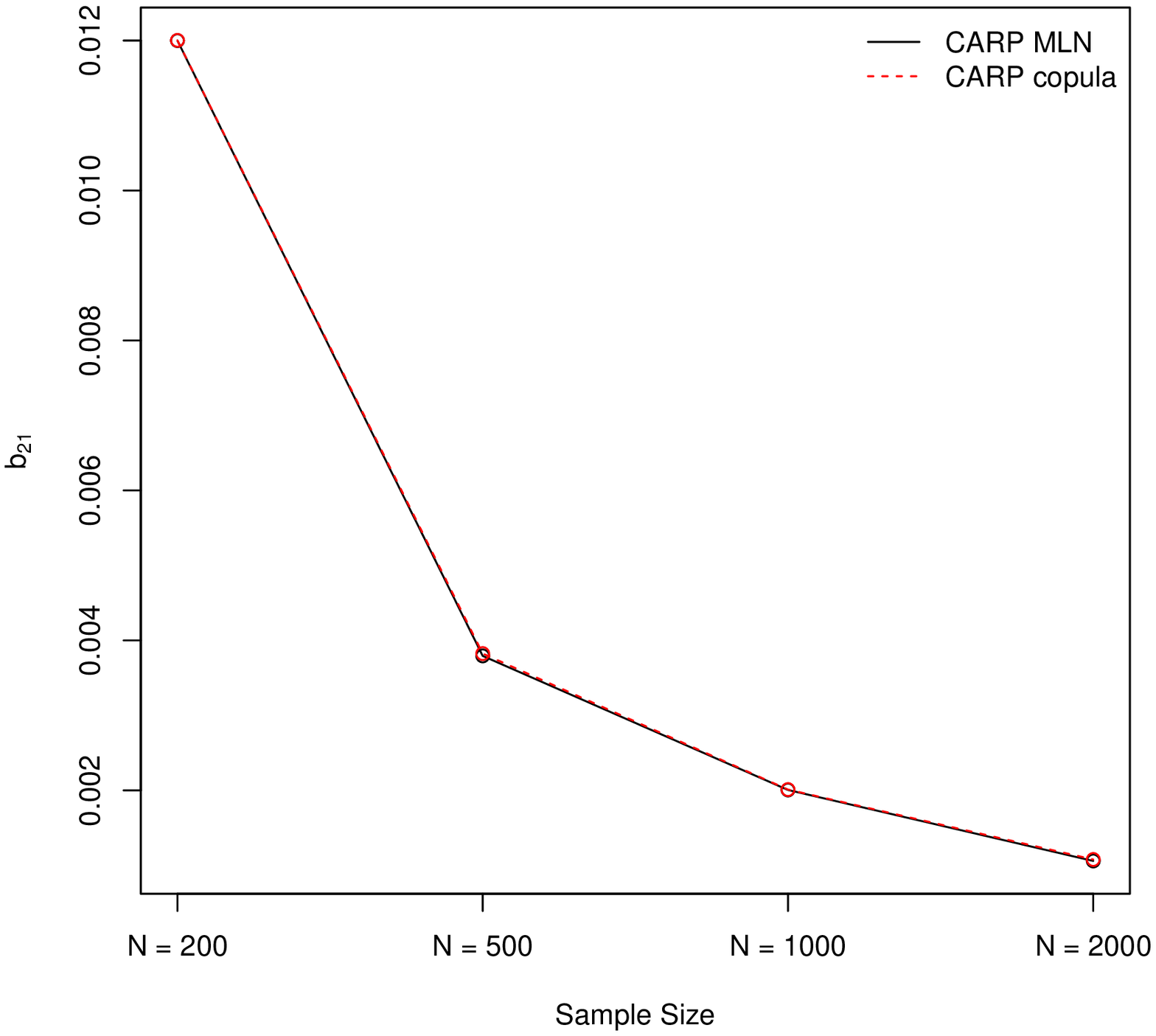}
		
	\end{subfigure}%
	~
	\begin{subfigure}[h]{0.5\textwidth}
		\centering
		\includegraphics[width=\textwidth, height=5.9cm]{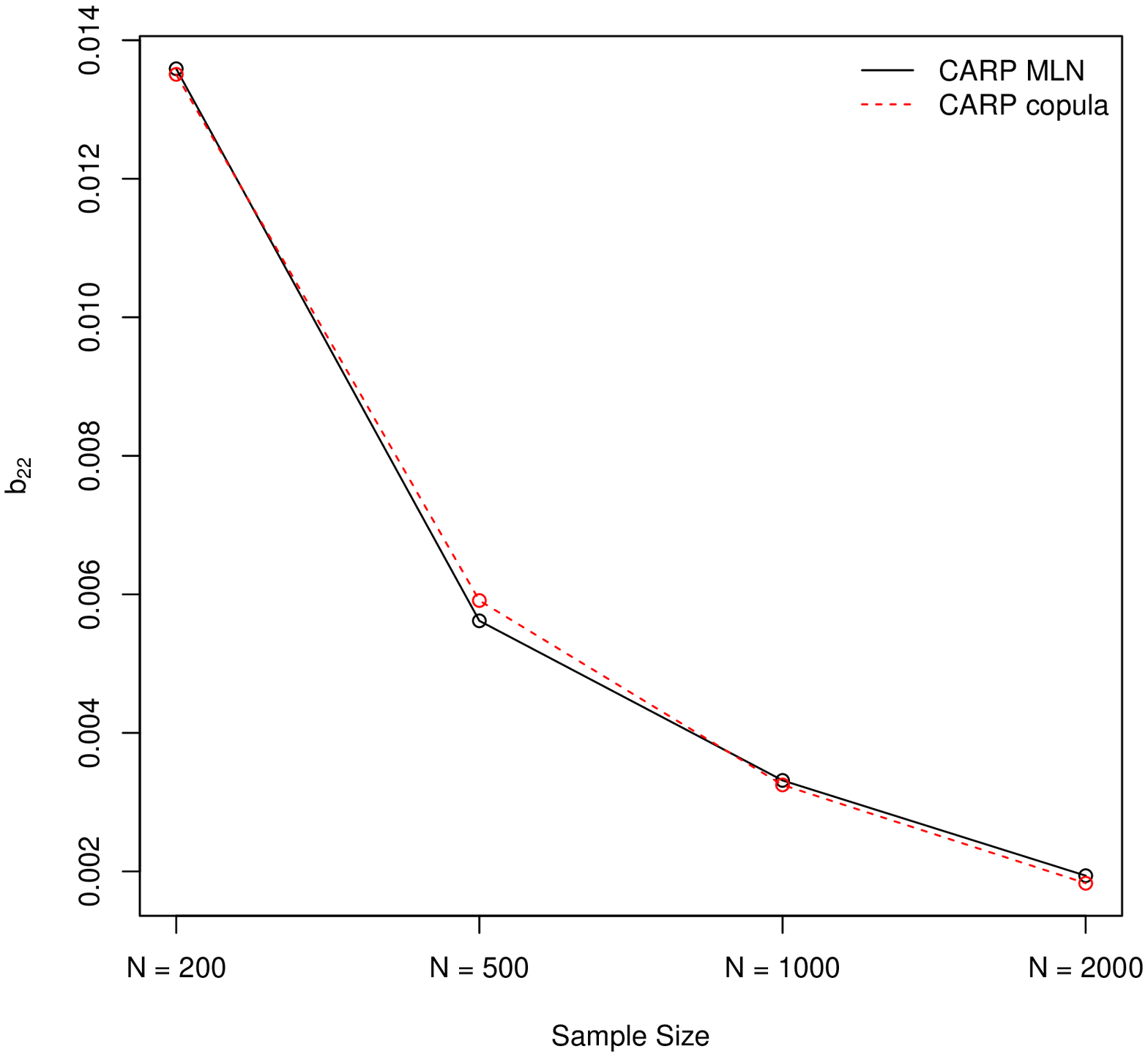}
		
	\end{subfigure}%
	~
	\caption{MSE for the location parameter $\mu_1$, $\mu_2$, scale parameter $\sigma_1$, $\sigma_2$ and $\Bvec$, calculated by both CARP-MLN and copula model with different sample sizes. The true model used to generated data is CARP-MLN.}
	\label{Fig:result2}
\end{figure}

Table~\ref{table:result1} shows the average AIC of the fitted models summarized over the 1000 simulated data sets generated using both the CARP-copula and CARP-MLN models at different sample sizes.  Both the CARP-MLN and CARP-copula models were applied to each simulated data set. We can see at each fixed sample size, the fitted CARP model that matches the true model used for simulating the data generally outperforms the other model by having a smaller average AIC value, except for the smallest size case at $n=200$ where the difference in the average AIC values are extremely small. The advantage of using the true model becomes more prominent as the sample size increases.

Figures \ref{Fig:result1} and \ref{Fig:result2} compare the MSE of the estimated parameters at different sample sizes between the two fitted models for the simulated data generated from the CARP-copula and the CARP-MLN models, respectively. The MSEs are evaluated for the location parameter $\mu$, the scale parameter $\sigma$ and linear transformation coefficient $\Bvec$ which are included in both figures. The black solid lines represent the MSE based on the fitted CARP-MLN model, and the red dashed lines represent the estimates from the fitted CARP-copula model. Two major conclusions can be drawn from these two figures. First, the MSEs of all the parameters decrease with the sample size. The more data we generate and use to fit the models, the more reliable the estimated models are with more accurate estimates of model parameters regardless of the choice of the model. Second, the use of the correct model does lead to slightly more accurate estimate of the model parameters across all sample size and model parameters. However, the improvement is more prominent for the scale parameters $\sigma$ than other model parameters.  When the true underlying model is generated by CARP-copula and far from bivariate lognormal, CARP-copula fits data better.

Table \ref{table:result2} shows the average AIC values of the fitted models when the Kendall's tau for capturing the dependence between two event types varies at the levels of 0, 0.11, 0.33 and 0.55. When the Kendall's tau is not zero which suggests some level of dependence between two marginal variables, using a fitted model that matches the true model for generating the data will result in a smaller average AIC value with a bigger improvement achieved when stronger dependence exists between the two types of events.  When the Kendall's tau is zero indicating the marginal distributions are independent, the average AIC values are generally similar regardless which model is used to fit the data.

Table \ref{table:result5} shows the comparison based on using different scale parameters in the true underlying model for both the CARP-copula and CARP-MLN models. Again, across all evaluated scale parameter values, we observe consistently smaller average AIC values when the fitted model matches the true model used for data generation.

Lastly, Table \ref{Table:misspecfication} compares the results of using different coefficient matrix $\Bvec$. We simulated data with both non-zero and zero $\Bvec$ using both the CARP-MLN and CARP-copula models. For each simulated data, we fitted both the CARP-MLN and CARP-copula models with non-zero and zero $\Bvec$. Similar patterns can be observed. When the true model was used as the fitted model, it results in the smallest AIC value. When non-zero $\Bvec$ was used to generate the data, the use of covariate adjustment led to substantial improvement in the average AIC of the fitted model compared to using zero $\Bvec$. On the other hand, when the data were generated with zero $\Bvec$, using the non-zero $\Bvec$ fitted models produced similar AIC values as the zero $\Bvec$ models. Therefore, the covariate adjustment is generally recommended due to its potential to substantially improving the model performance by leveraging the additional covariate information.

\section{Analysis of the Geyser Data}\label{sec:application}

In this section, we present the analysis for the bivariate geyser system in the Yellowstone National Park using the proposed CARP models. Two adjacent geysers including the West Triplet and the Grotto Geyser are considered for our analysis. Geyser eruptions are highly related to underground water levels, which can be affected by a nearby geyser eruption. Also, it is believed that the eruption duration could affect the gap time until the next eruption. In particular, the longer the current eruption lasts, the longer it will take for the next eruption to occur. This is because a longer eruption usually indicates more water consumption during the eruption and hence a longer water gathering time is expected to reach the next eruption.

We use both the CARP-MLN and the CARP-copula models with lognormal marginal distributions to analyze the geyser data. For the CARP-MLN model, the parameters are $\thetavec = ( \mu_{1}, \mu_{2}, \sigma_{1}, \sigma_{2}, \eta, b_{11}, b_{12}, b_{21}, b_{22})^\prime$, where $\mu_{1}, \mu_{2}, \sigma_{1}, \sigma_{2} \textrm{ and } \eta$ define the baseline location parameters and scale parameters in the bivariate lognormal distribution. For the CARP-Copula model, the parameters are $\thetavec = ( \mu_{1}, \mu_{2}, \sigma_{1}, \sigma_{2}, \alpha, b_{11}, b_{12}, b_{21}, b_{22} )^\prime$, where $\mu_{1}, \mu_{2}, \sigma_{1} \textrm{ and } \sigma_{2}$ are the location and scale parameters in marginal lognormal distributions, and $\alpha$ is the coefficient parameter in the Gumbel copula. In both cases, the 2 $\times$ 2 matrix $\Bvec$ is the linear coefficient, and it can be denoted as
$$\Bvec = \left(\begin{array}{cc} b_{11} & b_{12} \\ b_{21} & b_{22} \end{array}\right).$$

Table \ref{table:geyser_result_lognormal} shows estimation results for the CARP-MLN model, while Table \ref{table:geyser_result_copula} presents estimation results for the CARP-copula model. For simplicity, we use $\widehat{\thetavec}_{\textrm{MLN}} $ and  $\widehat{\thetavec}_{\textrm{CP}} $ to represent parameter estimates based on the CARP-MLN and CARP-copula models, respectively.

The estimated linear coefficients $\widehat{\Bvec}$ for both models are shown as follows,
\begin{align}\label{eqn:estimated_B}
\widehat{\Bvec}_{\textrm{MLN}} =
\left(\begin{array}{cc}
0.880 & -0.003 \\
0.058 & 0.063
\end{array}\right) \quad
\text{ and }\quad
\widehat{\Bvec}_{\textrm{CP}} =
\left(\begin{array}{cc}
0.862 & -0.017 \\
0.081 & 0.062
\end{array}\right).
\end{align}
The estimates in \eqref{eqn:estimated_B} indicate that in general the longer the previous eruption duration is, the longer waiting time it takes until the next eruption. The eruption duration impact is substantial for the West Triplet Geyser as shown by $\widehat{b}_{11}= 0.880$ in $\widehat{\Bvec}_{\textrm{MLN}}$, which suggests the marginal previous eruption duration effect of West Triplet Geyser on its location parameter is 0.880 on average.
\begin{table}
	\caption{Parameter estimates and 95\% confidence intervals from CARP-MLN model for the geyser data.}
	\label{table:geyser_result_lognormal}
	\begin{center}
		\begin{tabular}{ c | rrr}
			\hline
			\hline
			Parameter & Estimates & 95\% lower & 95\% upper\\
			\hline
			$\mu_{1}$ &  1.881 & 1.839 & 1.922 \\
		$\mu_{2}$ &  2.126 & 2.072 & 2.180 \\
		$b_{11}$   &  0.880 & 0.720 & 1.041 \\
		$b_{21}$   &  0.058 &$-$0.166 & 0.283 \\
		$b_{12}$   &  $-$0.003 & $-$0.008 & 0.003 \\
		$b_{22}$   & 0.063  & 0.050 & 0.077 \\
		$\eta$       & $-$0.053  & $-$0.093 & $-$0.014 \\
		$\sigma_{1}$ & 0.416 & 0.391 & 0.440 \\
		$\sigma_{2}$ & 0.493 & 0.459 & 0.527 \\
		$\tau$ & $-$0.069 & $-$0.070 & $-$0.067\\
			\hline
			\hline
		\end{tabular}
	\end{center}
\end{table}
\begin{table}
	\caption{Parameter estimates and the corresponding 95\% confidence intervals from CARP-copula model using lognormal marginal distributions for the geyser data.}
	\label{table:geyser_result_copula}
	\begin{center}
		\begin{tabular}{ c | rrr}
			\hline
			\hline
			Parameter & Estimates & 95\% lower & 95\% upper\\
			\hline
			$\mu_{1}$ &  1.847 & 1.809 & 1.886 \\
		$\mu_{2}$ &  2.102 & 2.051 & 2.152 \\
		$b_{11}$ & 0.862 & 0.700 &  1.024 \\
		$b_{21}$ & 0.081 &  $-$0.147 & 0.311 \\
		$b_{12}$ & $-$0.017 & $-$0.007 & 0.004 \\
		$b_{22}$ & 0.062 & 0.047 & 0.076  \\
		$\sigma_{1}$ & 0.417 & 0.392 & 0.442\\
		$\sigma_{2}$ & 0.497 & 0.459 & 0.535 \\
		$\alpha$ & 1.000 & 0.958 & 1.042\\
		$\tau$ & 0.000 & $-$0.001 & 0.001 \\
			\hline
			\hline
		\end{tabular}
	\end{center}
\end{table}

For the CARP-MLN model, the estimated covariance matrix is
$$\widehat{\Sigmavec} =
\left(\begin{array}{cc}
0.172 & -0.022  \\
-0.022 & 0.246
\end{array}\right),$$
where the correlation estimation is calculated as $\widehat{\eta} / \sqrt{\widehat{\sigma}^2_2 + \widehat{\eta}^2} = -0.05$. This indicates a small negative correlation between the event gap times of the West Triplet and Grotto Geysers. In other words, a longer eruption gap time for West Triplet Geyser could be associated with a shorter time interval for the next eruption of the Grotto Geyser. The Kendall's tau provides a measure on event dependence in both models. The estimates from the CARP-MLN and the CARP-copula models are $\widehat{\tau}_{\textrm{MLN}}=-0.07$ and $\widehat{\tau}_{\textrm{CP}} = 0$, respectively. The confidence intervals for $\tau$ is calculated by the Delta method in Appendix~\ref{Confidence Interval for Kendall's tauy}. The calculated AICs for the CARP-MLN and CARP-copula models are 5113.1 and 5120.5, respectively, indicating the CARP-MLN model is a slightly better fit for the geyser data.

\begin{figure}
	\centering
	\includegraphics[width=.6\linewidth, height=9.5cm]{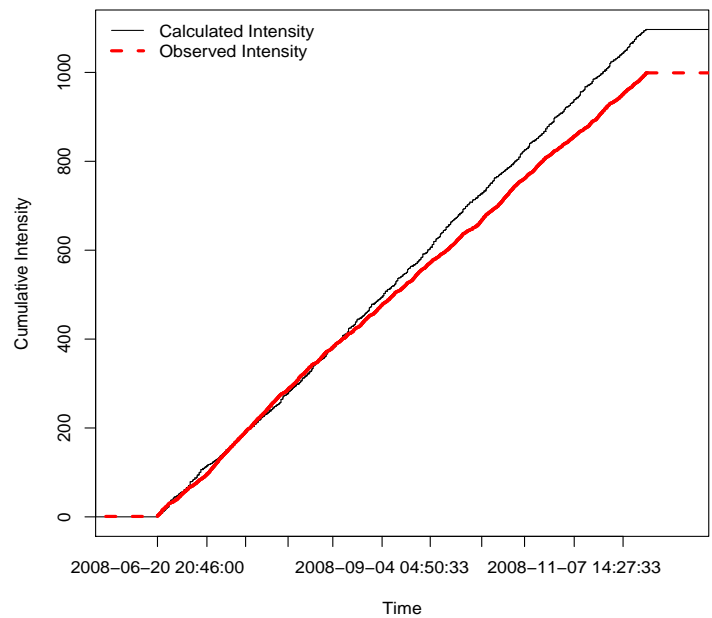}
	\captionof{figure}{Cumulative intensity function from fitted CARP-MLN model using the geyser data. }
	\label{fig:lognormal_intensity}
\end{figure}

\begin{figure}
	\centering
	\includegraphics[width=.6\linewidth, height=9.5cm]{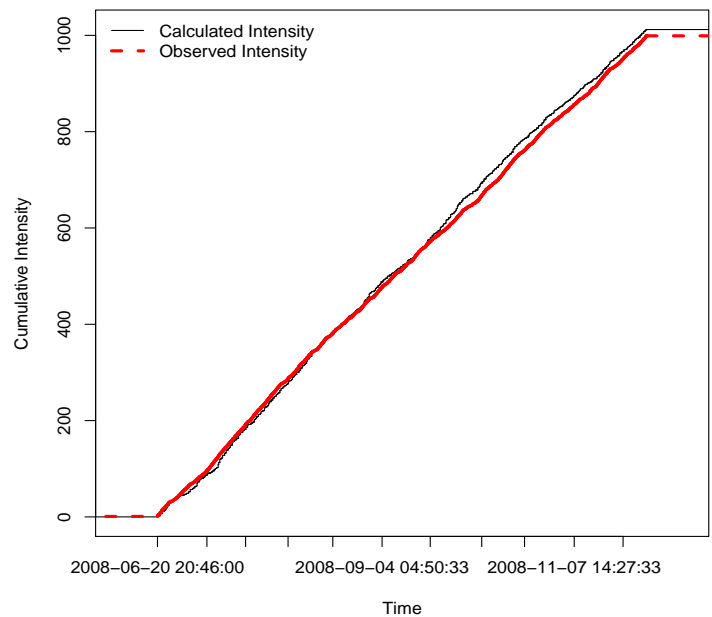}
	\captionof{figure}{Cumulative intensity function calculated from the CARP-copula model with the lognormal marginal distributions.}
	\label{fig:copula_intensity}
\end{figure}

One way to measure the goodness of fit is to compare the estimated cumulative intensity function with the observed. The estimated and observed cumulative intensity functions from both models are shown in Figures \ref{fig:lognormal_intensity} and \ref{fig:copula_intensity}, with the black solid lines and the red dashed lines representing the estimated and the observed cumulative intensity functions respectively. We can see a better agreement between the estimated and the observed cumulative intensity functions based on the CARP-copula model compared with the CARP-MLN model.

\section{Concluding Remarks}\label{sec:conclusion}
This paper introduces two CARP models, CARP-MLN and CARP-Copula, for modeling a bivariate recurrent process. With the covariate adjustment, the CARP models provide improvements over traditional models for capturing the interdependence of the bivariate event system. When the underlying data are close to a bivariate lognormal distribution, both models work similarly well. However, when the real data are inconsistent with a bivariate lognormal distribution, the CARP-copula model is recommended to improve the model performance.

A special case of Archimedean copulas, the Gumbel copula function, is considered in this paper. However, the general method can easily be adapted for other copula functions available for the chosen models. One example is the Frank copula. Compared to the Gumbel copula where only positive dependence can be quantified, the Frank copula can accommodate both positive and negative dependence between two event types. On the other hand, the Gumbel copula has an asymmetric dependence structure where correlations on the tail can be very different than both the Gaussian and Frank copulas. In addition, differences of the joint density functions between the Frank and Gaussian copulas are negligible when the marginal distributions are the same.

The adjustment by using effective covariates significantly improves the model performance. We have shown through the simulation study, when the true model includes a significant covariate effect, that using the covariate adjustment significantly improves model fitting. On the other hand, if there is no significant covariate effect, using the covariate adjustment does not raise the AIC values. Therefore, in general we recommend using the covariate adjustment given the true underlying model is unknown.

The choice of the marginal distributions in the CARP-copula model was made based on the AIC value. Marginal distributions achieving the minimum AIC value were selected in our model. One advantage of using the CARP-copula model is that different marginal distributions can be easily used to model different event types. This offers tremendous flexibility and broader generality to the CARP-copula model.

A multivariate CARP model will be considered for future work when there are more than two event types in the recurrent system. For the geyser application, other covariates in addition to the eruption duration would also be explored to further improve the performance of the CARP models. In engineering applications when testing the reliability of systems is often of interest. The loading-sharing system is related to the CARP model (e.g., \citealp{smith1983limit}, \citealp{tierney1982asymptotic}, \citealp{sutar2014accelerated},  and \citealp{zhang2020reliability}). In the future, it will be interesting to apply the CARP to model the reliability of loading sharing systems.

\section*{Acknowledgments}
The authors thank the editor, associate editor, and two referees, for their valuable comments that helped in improving the paper significantly. The authors acknowledge the Advanced Research Computing program at Virginia Tech for providing computational resources. The work by Hong was partially supported by National Science Foundation Grant CMMI-1904165 to Virginia Tech.

\appendix
\section{Appendix}
\subsection{Conditional Lognormal Probability \label{Conditional Lognormal Probability}} 

If a bivariate random variable $\yvec = (y_{1}, y_{2})'$ follows a lognormal distribution $\textrm{MLN}(\muvec, \Sigmavec)$ where
$$
\muvec = (\mu_{1}, \mu_{2})' \quad
\textrm{ and }\quad
\Sigmavec = \left(\begin{array}{cc} \sigma_{11} & \sigma_{12}\\ \sigma_{21} & \sigma_{22} \end{array}\right),
$$
then the conditional distribution of $\log(y_{1}) | \log(y_{2})$ follows a normal distribution $\textrm{N}(\mu_{c}, \sigma_{c})$ with location and scale parameter as
$$
\mu_{c} = \mu_{1} +  \sigma_{12}\sigma^{-1}_{22} [\log(y_{2})- \mu_{2}]\quad
\text{ and }\quad
\sigma_{c} = \sigma_{11} - \sigma_{12}\sigma^{-1}_{22}\sigma_{21},
$$
respectively.

\subsection{Confidence Interval for Kendall's tau\label{Confidence Interval for Kendall's tauy}}
For lognormal cases, the Kendall's tau estimator can be expressed as
$$\widehat{\tau} = \frac{2}{\pi}\arcsin\left(\frac{\widehat{\eta}}{\sqrt{\widehat{\sigma}^2_2 + \widehat{\eta}^2}}\right),$$
where the asymptotic distribution is known from the ML estimator. Using the Delta method,
$$\textrm{Var}(\widehat{\tau}) = \left( \frac{\partial \widehat{\tau}}{\partial \widehat{\eta}}\right)^2 \textrm{Var}(\widehat{\eta}) + \left( \frac{\partial \widehat{\tau}}{\partial \widehat{\sigma}_2}\right)^2 \textrm{Var}(\widehat{\sigma}_2) + 2 \left( \frac{\partial \widehat{\tau}}{\partial \widehat{\eta}}\right) \left( \frac{\partial \widehat{\tau}}{\partial \widehat{\sigma}_2}\right) \textrm{cov}(\widehat{\eta}, \widehat{\sigma}_2),$$
where
$$ \frac{\partial \widehat{\tau}}{\partial \widehat{\eta}} = \frac{2}{\pi} \ffrac{1}{\sqrt{1 - \frac{\widehat{\eta}^2}{\widehat{\sigma}^2_2 + \widehat{\eta}^2} }} \left[ (\widehat{\eta}^2 + \widehat{\sigma}^2_2)^{-\frac{1}{2}} - \frac{1}{2} \widehat{\eta} (\widehat{\eta}^2 + \widehat{\sigma}^2_2)^{-\frac{3}{2}}  \right], $$
and
$$ \frac{\partial \widehat{\tau}}{\partial \widehat{\sigma}_2} = -\frac{2}{\pi} \ffrac{1}{\sqrt{1 - \frac{\widehat{\eta}^2}{\widehat{\sigma}^2_2 + \widehat{\eta}^2} }} \left[ - \widehat{\eta}\widehat{\sigma}_2 (\widehat{\eta}^2 + \widehat{\sigma}^2_2)^{-\frac{3}{2}} \right].$$
For the Gumbel copula case,
$$\widehat{\tau} = 1 - \frac{1}{\widehat{\alpha}}.$$
Similarly with the lognormal case, the variance for the estimator can be written by using the Delta method. That is,
$$\Var(\widehat{\tau}) = \Big[ \frac{\partial g(\widehat{\alpha})}{\partial \widehat{\alpha}} \Big]^2 \Var(\widehat{\alpha}),$$
where $g(x) = 1 - 1/x.$




\end{document}